\documentclass[fleqn,usenatbib]{rasti}

\usepackage{newtxtext,newtxmath}

\usepackage[T1]{fontenc}

\DeclareRobustCommand{\VAN}[3]{#2}
\let\VANthebibliography\thebibliography
\def\thebibliography{\DeclareRobustCommand{\VAN}[3]{##3}\VANthebibliography}

\usepackage{booktabs}

\usepackage{graphicx}	%
\usepackage{placeins}	%
\usepackage{amsmath}	%
\usepackage{cleveref}
\usepackage{chemformula}
\usepackage{amsmath}	%
\usepackage{tikz}
\usepackage{multirow}

\usepackage{listings}
\usepackage{xcolor}

\usepackage{tikz}
\usetikzlibrary{arrows.meta, calc, positioning, fit, backgrounds}

\definecolor{pyfill}{RGB}{233,239,248}
\definecolor{pyline}{RGB}{47,95,168}
\definecolor{fofill}{RGB}{230,242,239}
\definecolor{foline}{RGB}{31,122,108}
\definecolor{artfill}{RGB}{246,236,221}
\definecolor{artline}{RGB}{165,102,31}
\definecolor{bandfill}{RGB}{247,247,245}
\definecolor{bandline}{RGB}{225,225,222}
\definecolor{muted}{RGB}{110,110,110}

\newcommand{\subtext}[1]{\\[3pt]{\scriptsize\ttfamily\selectfont\color{muted}#1}}
\newcommand{\note}[1]{\\[3pt]{\scriptsize\itshape\selectfont\color{muted}#1}}

\title[\texttt{UCLCHEM} 4.0]{\texttt{UCLCHEM} 4.0: An open source gas-grain astrochemistry simulation framework}

\author[G. Vermari\"en et al]{%
Gijs Vermari\"en$^{1}$\thanks{E-mail: vermarien@strw.leidenuniv.nl},
Serena Viti$^{1,2,3}$, 
Tobias~M.~Dijkhuis$^{1,4,5}$,
Le Ngoc Tram$^{1}$,
Marcus Keil$^{6}$,
\newauthor
Katarzyna~M.~Dutkowska$^{1}$,
Felix~D.~Priestley$^{7}$.
\\
$^{1}$Leiden Observatory, Leiden University, PO Box 9513, 2300 RA Leiden, The Netherlands
\\
$^{2}$Transdisciplinary Research Area (TRA) `Matter'/Argelander-Institut für Astronomie, University of Bonn, 53121 Bonn, Germany
\\
$^{3}$Department of Physics and Astronomy, University College London, Gower Street, London, UK
\\
$^{4}$Leiden Institute of Chemistry, Gorlaeus Laboratories, Leiden University, PO Box 9502, 2300 RA, Leiden, The Netherlands
\\
$^{5}$Institute for Molecules and Materials, Radboud University, 6525 AJ Nijmegen, The Netherlands \\
$^{6}$ Department of Computer Science, University College London, London, United Kingdom\\
$^{7}$ School of Physics and Astronomy, Cardiff University, Queen’s Buildings, The Parade, Cardiff CF24 3AA, UK
}
\date{Accepted XXX. Received YYY; in original form ZZZ}

\pubyear{\the\year{}}

\begin{document}
\label{firstpage}
\pagerange{\pageref{firstpage}--\pageref{lastpage}}
\maketitle

\begin{abstract}
Astrochemical modelling is a key tool for interpreting the formation and 
destruction of molecules in the interstellar medium, complementing the increasingly detailed observations enabled by modern day observational facilities.
\texttt{UCLCHEM} is a comprehensive gas-grain astrochemical modelling framework that can compute the chemical evolution inside a parcel of gas and dust under a specified, time-dependent physical description. This allows both the post-processing of hydrodynamical simulations and the usage of physical parametrizations, ranging in context from extra-galactic to protoplanetary disk scales.
The framework consists of a core routine that solves gas-grain chemical reaction networks as a function of time, self-consistently coupled to either
a user-defined or parametrized physical evolution of density, temperature and other physical input parameters.

The chemistry is solved with a rate-equation formalism, and includes a three phase description, namely gas phase, ice surface and bulk ice. This includes transport between the gas and ice phases via freeze-out, thermal and non-thermal desorption mechanisms, as well as a geometrical treatment of the growth and shrinkage of the surface and bulk chemistry.
The framework
includes parametrizations for astrochemical modelling of cloud collapses, protostellar cores and shocks as well as the ability to provide user-defined inputs.
 This release of the framework focuses on many quality-of-life-improvements, including an object-oriented Python interface, an upgrade from two phase to three phases and expanded support for post-processing of hydrodynamical simulations and user-defined parametrizations.
 This manuscript provides an overview of the physics and chemistry included in \texttt{UCLCHEM}, as well as the inner workings of the solver routine and the programming interface.
\end{abstract}


\section{Introduction}
The Interstellar Medium (ISM) spans a large range of densities and temperatures, leading to steep changes in the energetics and dynamics of the gas. Chemical modelling is therefore essential to predict the atomic and molecular abundances as a function of such changes. Chemical models have increased in complexity over the years, due partly to the need to be able to match more  accurate and spatially resolved observations, and partly due to our increase in the knowledge of the chemical processes that occur in the gas as well as on the dust grains. 
In the last 20-30 years time dependent gas-grain chemical models have been developed by several teams. In most of them, rate equations are solved to determine the rate of change in chemical species. Each code is more or less complicated depending on the purpose that they were built for. For example, chemical models that were originally constructed to be part of a hydrodynamical framework tend to solve a simplified reduced chemistry, often with the dust having a very minor chemical role (e.g.~acts as a catalyst to form molecular hydrogen only) 
(e.g.~ \citealt{peterssonNOCTUASuiteSimulations2025}; \citealt{khatriHYACINTHHYdrogenCarbon2024})
while codes that have developed to explain the chemical complexity in space will sacrifice the complexity of the physics in favour of the inclusion of large chemical networks leading to the formation of complex organic molecules (COMs) 
(e.g.~\citealt{borshchevaFormationComplexOrganic2025}; \citealt{quenardChemicalModellingComplex2018})

As the simulation of molecules spans
levels of detail and has to be consistent through
different astrophysical stages of evolution, 
tailored astrochemical codes exist for each purpose. Each of these includes
different micro and macro physics and chemistry, 
with different simulation methods. 
One such group is rate equation astrochemical codes that solve a differential equation tracing different phases
of species, often including gas, grain ice surfaces and bulk ices. 
These codes describe a chemical network modelling the interstellar medium as a system of differential equations (ODE),
with each species corresponding to an ODE, and
each reaction corresponding to a differential equation term, often grouped by chemical reaction mechanisms. Important codes in this group, which
\texttt{UCLCHEM} belongs to as well, are 
\texttt{NAUTILUS} \citep{ruaudGasGrainChemical2016}, \texttt{Magickal} \citep{garrod2013} and \texttt{MONACO} \citep{vasyuninFormationComplexMolecules2017}. 
Alternatively, instead of formulating the chemistry using a rate equation,
the chemistry can be modelled as a stochastic system using Kinetic Monte
Carlo simulations, as described and benchmarked against the last 4 codes
in \citet{jimenez-serraModellingMethanolHydride2025}. A third approach, the direct master equation method
\citep{stantchevaDeuteriumFractionationInterstellar2003}, instead solves for the full probability
distribution over the discrete population states of low-abundance,
highly reactive surface species, while retaining the ODEs
for the remaining (gas phase) network. This avoids the stochastic noise inherent to
Monte Carlo methods, but its computational cost grows rapidly with the
number of species treated this way, restricting its use to targeted
studies of specific reactive species rather than full chemical networks. A more domain 
specific type of code is the development of codes for
the simulation of disks. These codes potentially include heating and cooling via thermochemistry, and have to deal with the large dynamical range of temperatures, radiation and density. Examples include ALCHEMIC \citep{semenovCHEMICALEVOLUTIONTURBULENT2011}, DALI \citep{brudererSurvivalMolecularGas2013}, PRIZMO \cite{grassiModellingThermochemicalProcesses2020} and ROKKO \citep{furuyaWaterDeuterationOrthotopara2015, benedettiCODiffusionInterstellar2025}. 
Another class of codes, that instead focuses on
less dense interstellar environments with radiation and thermochemistry, are photodissociation region (PDR) codes. Many codes for simulating these environments exist, including sibling codes
of \texttt{UCLCHEM}: \texttt{UCLPDR} \citep{bellChemistryTransientMicrostructure2005}, which is no longer actively maintained, and \texttt{3D-PDR} \citep{bisbas3DPDRNewThreedimensional2012}, which is actively maintained and improved \citep{zhuRAYTHEIAHighperformanceRaytracing2026}. Efforts to benchmark the many existing PDR codes were performed in \citet{rolligPhotonDominatedRegion2007}, and still function as a reference point for new PDR codes. Another more 
general framework for (thermo)chemistry exists,
namely \texttt{KROME} \citep{grassiKROMEPackageEmbed2014}, which not only has the goal of standalone chemical simulations, but also provides the
ability to integrate astrochemistry into hydrodynamical simulations in Fortran.
\texttt{UCLCHEM} is a time dependent gas-grain chemical model that has been developed and augmented over many years and is, since 2017, in an open source format (\url{https://uclchem.github.io/}). 
It was first developed in 1999 with the aim of modelling the clumps ahead of Herbig-Haro objects \citep{vitiChemicalEvolutionAhead1999} and the enhanced molecular emission of hot cores \citep{vitiTimedependentEvaporationIcy1999} 
and underwent various major chemical updates; for example, time dependent thermal desorption \citep{vitiEvaporationIcesMassive2004}, non-thermal desorption\citep{robertsDesorptionInterstellarIces2007}, and a C-shock treatment \citep{ vitiL1157B1WaterAmmonia2011}
based on \citep{jimenez-serraParametrizationCshocksEvolution2008} parametrization. Its previous release was in 2017 \citep{holdshipUCLCHEMGasgrainChemical2017}.
This work presents the newest official release of \texttt{UCLCHEM} and describes  all the changes and augmentation performed to the code since 2017. We note that some changes have been implicitly presented over the years in scientific papers and we shall refer to these for the details of the implementations when appropriate. 

\begin{figure}
    \centering
    \includegraphics[width=1.0\linewidth]{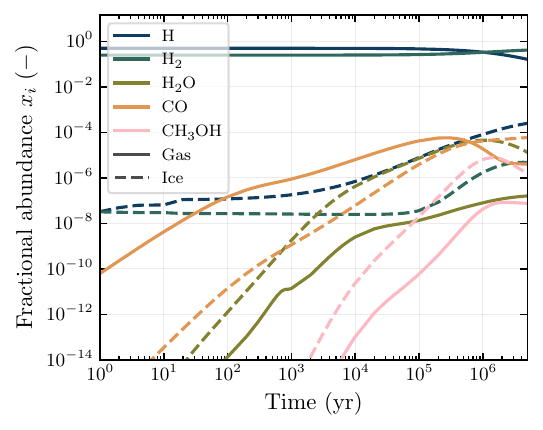}
    \caption{An example of the evolution of chemical species with \texttt{UCLCHEM}. As the
    system of differential equations is solved over time, some molecules are formed, whilst others are 
    destroyed.}
    \label{fig:example}
\end{figure}

\texttt{UCLCHEM} models the evolution of the abundance of different species in the gas phase and on the ices on
 dust grains in the ISM. In order to simulate chemistry, one must define a reaction network with different molecular species, typically indexed by
  $i$, and reactions between the species, typically indexed by $j$, covering the creation and destruction pathways of species of interest. To each species we can attribute a number density $n_i\;(\rm cm^{-3})$ and hence a fractional abundance with respect to the hydrogen nuclei number density: $x_i = n_i / n_\mathrm{H,nuclei}$, where
  $n_\mathrm{H,nuclei}=n_{\ch{H}}+2n_{\ch{H2}}$. Each reaction then has a rate constant and equation associated with it: $k$, determining the speed at which the reaction occurs together with the densities of the reactants involved. This means that for each species we can then write down a differential equation that accounts for production and destruction reactions of species:
\begin{equation}
\frac{\mathrm{d} n_i(t)}{\mathrm{d} t}=\sum_j\pm k_{j} n_{j_1} n_{j_2}+\sum_j \pm k_j n_j,
\end{equation}
with the positive terms being production and the negative ones being destruction through bi- and uni-molecular reactions respectively, and $n_j$, $n_{j_1}$ and $n_{j_2}$ are the number densities of the reactants
involved in each reaction $j$.
If we then use a numerical differential equation solver, we can obtain the trajectory of each species 
over time, $n_i(t)$. 
We provide an example of a static isothermal cloud evolving from its initial 
conditions in \Cref{fig:example}. 
The \texttt{UCLCHEM} framework provides easy and interoperable modules 
that can build more complex sequential models, grids of models for parameter studies and
post-processing of hydrodynamical simulations, but the core solver routine remains the same. \texttt{UCLCHEM} also contains extensive routines that assist the user with building reaction
networks, which are combined into a \texttt{Makerates} submodule.

In this manuscript, we provide a broad overview of the chemical and physical modules of \texttt{UCLCHEM},  as well as its current Fortran and Python implementation. We shall illustrate the important modules via figures highlighting common workflows and effects.
In this article we create an arbitrary division between the treatment of the physics and the chemistry. 
Even though the two are nonlinearly coupled and tightly integrated, this division reflects the division
of the underlying code. We first discuss the chemistry in \Cref{sec:chem}, then continue with the discussion of
the physics in \Cref{sec:phys}. We then describe the workings of the simulation framework itself in \Cref{sec:uclchem} and
lastly conclude the manuscript with \Cref{sec:conclusion}.

\section{The chemistry}
\label{sec:chem}
As the simulation framework can deal with models of many different 
astrophysical environments, all chemical processes occurring in these environments need to be modelled. To this end, \texttt{UCLCHEM} includes
a treatment of gas phase and ice phase chemistry, both with their own
unique set of reaction mechanisms and physical parameters. The phases 
interact with one another via either the deposition of species onto the 
grains, or via desorption of species into the gas phase. A default 
assumption of \texttt{UCLCHEM} is that the dust and gas temperature
are coupled with each other, i.e.~$T_{\mathrm{dust}}=T_{\mathrm{gas}}$, unless specified otherwise. 

The treatment of the species on the grains includes a separate treatment of both the surface and the
bulk \citep{hasegawaModelsGasGrainChemistry1992, garrodFormationCO2Other2011, ruaudGasGrainChemical2016}. The surface is the outermost layer of ice, which can 
interact more freely among themselves and react with species in the gas phase. The bulk layers 
are the enclosed layers, which allow for less reactivity. Together we refer to these species captured
on the dust grains as ice or grain species. In \texttt{UCLCHEM} the symbols \#, @ and \$  are used to represent surface, bulk and ice species respectively.

We now provide an overview of the many mechanisms present to model the gas, surface, and bulk and their chemical interactions.

\subsection{Gas phase chemistry}
\begin{figure*}
    \centering
    \includegraphics[width=1.0\linewidth]{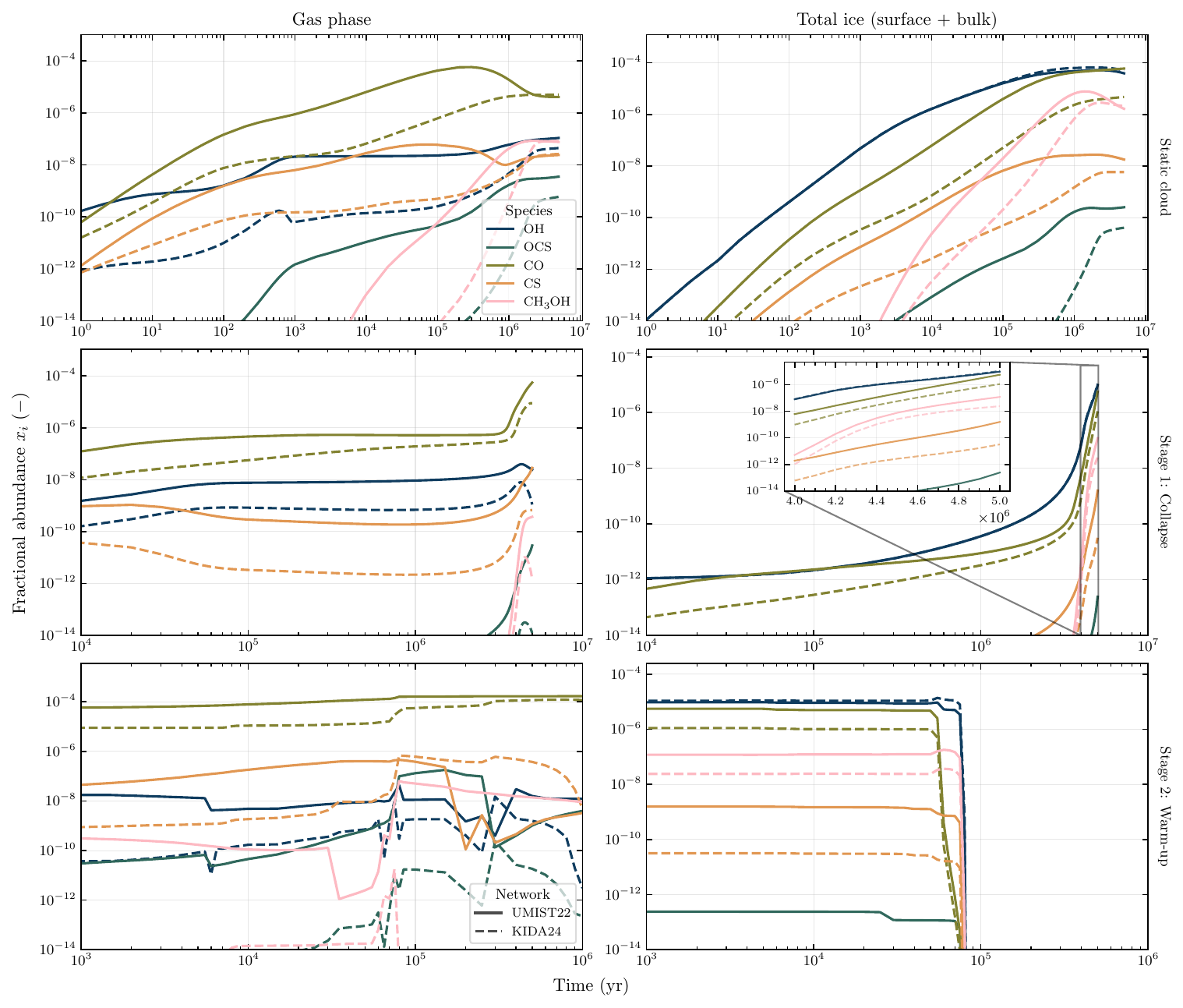}
    \caption{A comparison of the chemistry with the UMIST and KIDA gas phase reaction databases, for an isothermal cloud at constant density on the top row, a collapse model as stage 1 on the middle row and a protostellar model as stage 2 on the bottom row.}
    \label{fig:umistvkida}
\end{figure*}

In order to effectively simulate the gas phase chemistry of the interstellar medium, one
has to rely on estimates of the reaction rates from many different sources:
experiments, simulation, theory, and expert intuition. These databases aggregate many different reaction mechanisms between species, ionized species, cosmic rays, and radiation. 
\texttt{UCLCHEM} adopts a gas-phase
reaction agnostic approach where we can load different sets of gas-phase reactions. 
\texttt{UCLCHEM} relies on two main sources for this: The UMIST and KIDA databases \citep{millarUMISTDatabaseAstrochemistry2024, wakelam2024KIDANetwork2024}. A brief comparison of these two different databases for a 
static model at $n_\mathrm{H,nuclei}=10^{4}$~cm$^{-3}$ and $T=10$~K, a free-fall collapse model 
with a final density $n_\mathrm{H,nuclei}=10^{5}$~cm$^{-3}$ and a subsequent protostellar warm-up model with
a final temperature $T=300$~K is shown in \Cref{fig:umistvkida}. This figure shows similar patterns for many molecules, but with different exact solution trajectory due to the different chemical networks
with corresponding rate equations and reaction pathways. An in-depth analysis
of the differences between the respective chemical reactions and networks is beyond
the scope of this work.

At the start of each simulation, by default, we initialize the abundances in the gas
phase using the heavily depleted scenario from \citet{jenkinsUNIFIEDREPRESENTATIONGASPHASE2009}, but the user is free to choose initial abundances as a free parameter. Alternatively, for each model, the user
can choose to inherit the abundances from (the end of) a prior stage, or provide
their own abundances as starting values. Support for other elements and isotopes (other than those listed in \Cref{tab:jenkins})
is present, but their initial abundances are set to zero because these
are not covered within the default reaction network.

\begin{table}
\centering
\caption{Default initial elemental abundances relative to total H nuclei, from 
\citet{jenkinsUNIFIEDREPRESENTATIONGASPHASE2009} (heavily depleted case, Table~4). }
\begin{tabular}{lll}
\hline
Element & Parameter & Initial relative abundance $(-)$ \\
\hline
H          & $f_\mathrm{H}$ & $5.00 \times 10^{-1}$ \\
H$_2$      & - & $2.50 \times 10^{-1}$ \\
He         & $f_\mathrm{He}$ & $1.00 \times 10^{-1}$ \\
O          & $f_\mathrm{O}$ & $3.34 \times 10^{-4}$ \\
C          & $f_\mathrm{C}$ & $1.77 \times 10^{-4}$ \\
N          & $f_\mathrm{N}$ & $6.18 \times 10^{-5}$ \\
S          & $f_\mathrm{S}$ & $3.51 \times 10^{-6}$ \\
Mg         & $f_\mathrm{Mg}$ & $2.26 \times 10^{-6}$ \\
Fe         & $f_\mathrm{Fe}$ & $2.01 \times 10^{-7}$ \\
Si         & $f_\mathrm{Si}$ & $1.78 \times 10^{-6}$ \\
P          & $f_\mathrm{P}$ & $7.78 \times 10^{-8}$ \\
Cl         & $f_\mathrm{Cl}$ & $3.39 \times 10^{-8}$ \\
\hline
\end{tabular}
\label{tab:jenkins}
\end{table}

Unless stated otherwise, temperatures are given in K, number densities in cm$^{-3}$, and constants are given in cgs units.

The rate coefficient of each reaction between two gas-phase bodies are then described by the Arrhenius-Kooij equation:
\begin{equation}
k = \alpha \left(\frac{T_{\mathrm{gas}}}{300\ \mathrm{K}}\right)^\beta \mathrm{e}^{-\gamma / T_\mathrm{gas}},
\end{equation}
where $\alpha$  $(\mathrm{cm}^3\;\mathrm{s}^{-1})$, and $\beta$ (-) and $\gamma$ $(\mathrm{K})$ are parameters that can be found in aforementioned reaction databases, giving the rate coefficient for a bimolecular
reaction in  $\mathrm{cm}^3\;\mathrm{s}^{-1}$.
Another reaction type, this time unimolecular, is the reaction induced by cosmic ray ionization:
\begin{equation}
k = \alpha \zeta,
\end{equation}
with $\zeta$ the Cosmic Ray Ionization Rate (CRIR) in $\mathrm{s}^{-1}$, and $\alpha$ the dimensionless rate coefficient, so that $k$ is in $\mathrm{s}^{-1}$. For convenience, we quote specific values of $\zeta$ throughout this work as multiples of the reference value $\zeta_0=1.3\times10^{-17}\;\mathrm{s}^{-1}$ (e.g.\ $\zeta=10^5\zeta_0$), which avoids repeatedly writing out powers of ten; this is purely a shorthand for reporting numbers and $\zeta_0$ plays no further role in any formula. Cosmic rays can also induce reactions via secondary photons:
\begin{equation}
    k = \alpha \left(\frac{T}{300\ \mathrm{K}}\right)^\beta \frac{E}{1-\omega} \zeta,
\end{equation}
where $E$ is the (dimensionless) cosmic ray ionization efficiency, $\omega$ is the (dimensionless) dust grain albedo, and $\alpha$ is dimensionless. We also include
reactions driven by ultraviolet photons, with rate constant
\begin{equation}
    k = \alpha F_\mathrm{UV} \mathrm{e}^{-\gamma A_\mathrm{V}},
\end{equation}
with $\alpha$ in s$^{-1}$, $\gamma$ dimensionless, $A_\mathrm{V}$ the visual extinction in magnitudes, and $F_\mathrm{UV}$ the strength of the FUV radiation field in Habing units (dimensionless; see \Cref{sec:1drad} for its attenuation with depth).

We implement and include by default grain-assisted recombination for positively-charged ions of \ch{H}, \ch{He}, \ch{C}, \ch{Na}, \ch{Mg}, \ch{Si}, \ch{S}, and \ch{Fe}, which becomes important for cosmic ray ionization rates above $\sim$10$^{-16}$~s$^{-1}$ \citep{gongSimpleAccurateNetwork2017}. Ions of more electronegative elements, such as oxygen, typically recombine much more rapidly via charge transfer reactions. Following \citet{weingartnerElectronIonRecombinationGrains2001}, the grain-assisted recombination rate for ion $i$ with abundance $n_i$ is given by
\begin{equation}
    \frac{\mathrm{d}n_i}{\mathrm{d}t} = - k(T_{\rm gas}, \psi) \, n_i \, n_{\rm H,nuclei},
\end{equation}
where $k$ is in cm$^3$~s$^{-1}$ and
\begin{equation}
\psi=F_\mathrm{UV}\sqrt{T_{\rm gas}} / n_e,
\end{equation}
and the rate constant $k$ (in cm$^3$~s$^{-1}$, with the leading constant $0.6\times10^{-14}\,\mathrm{cm^3\,s^{-1}}$) is parametrized as
\begin{equation}
k(T_{\rm gas}, \psi) = \frac{0.6 \times 10^{-14}\,\mathrm{cm^3\,s^{-1}}\,{C_0}}{1+C_1\psi^{C_2}\left(1+C_3 T_{\rm gas}^{C_4}\psi^{-C_5-C_6\ln T_{gas}}\right)},
\end{equation}
with the coefficients $C_0-C_6$ listed in \Cref{tab:gar}. The factor of $0.6$ was adopted by \citet{gongSimpleAccurateNetwork2017} to better match the recombination rates determined observationally by \citet{wolfireChemicalRatesSmall2008}, and we impose a minimum $\psi$ value of $100$~K$^{1/2}$~cm$^3$ to avoid extrapolating beyond the range of validity of the parametrisation \citep{hunterImpactGMCCollisions2023}. 
\begin{figure}
    \centering
    \includegraphics[width=1.0\linewidth]{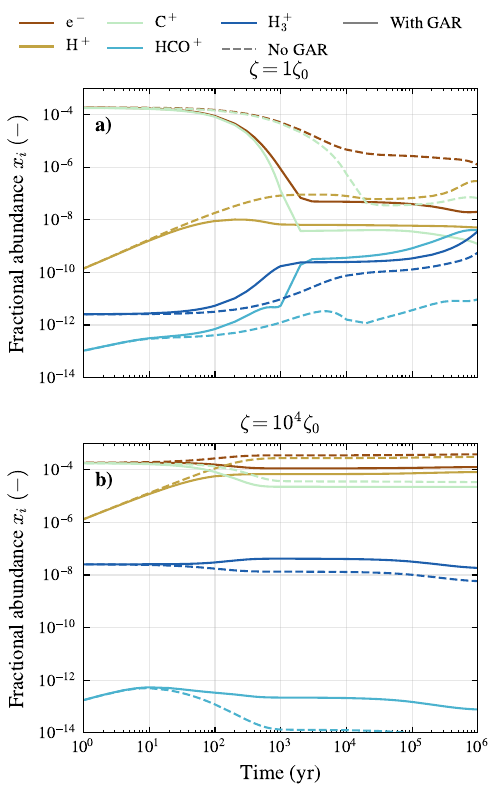}
    \caption{A comparison between a model with and without Grain Assisted Recombination (GAR). The model was run for normal and high $\zeta$ with model parameters $n_{\ch{H},\mathrm{nuclei}}=10^{4}$~cm$^{-3}$, $T=10$~K and $A_\mathrm{V}=12$.}
    \label{fig:gar-comparison}
\end{figure}

This treatment of grain-assisted recombination does not conserve charge, as positive ions are neutralised without also removing free electrons or modifying the grain charge distribution (not modelled in \texttt{UCLCHEM}). We therefore set the electron abundance to the sum of all the positively-ionised species to ensure charge conservation, i.e.~$n_e = \sum_{i\in \mathrm{ions}} n_i$, with the corresponding rate equation $\frac{\mathrm{d}n_e}{\mathrm{d}t} = \sum_{i\in \mathrm{ions}} \frac{\mathrm{d}n_i}{\mathrm{d}t}$. The effect of the
grain-assisted recombination for both default and high
CRIR is shown in \Cref{fig:gar-comparison}.

\subsection{Astrochemistry on the grains}
\label{sec:astrochem_on_grain}
One of the main features of \texttt{UCLCHEM} is treating chemistry in both the gas and ice phase.
At the start of a new model, species are initialized in the gas phase, so the ice abundances are zero (unless custom initial abundances are supplied).
Given a low enough temperature, the species can freeze out onto the grains, react with other species on the grains and then be desorbed back into the gas phase.
\subsubsection{Moving species between the gas, surface and bulk.}
\label{sec:surfbulkinteraction}
\begin{figure}
    \centering
    \includegraphics[width=1.0\linewidth]{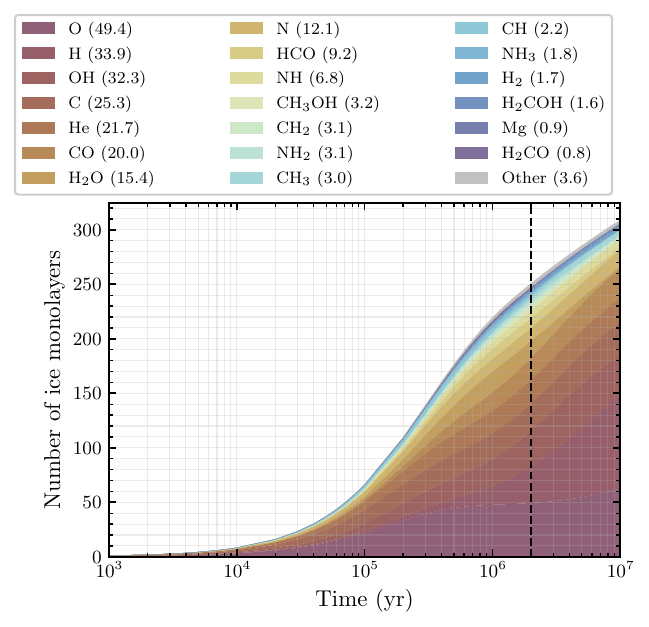}
    \caption{The formation of the ice in monolayers at $T=10$~K and $n_\mathrm{H}=10^{4}$~cm$^{-3}$. The number of monolayers at the time $t=2 \times10^6$~years is highlighted in the parentheses in the legend.}
    \label{fig:monolayers}
\end{figure}

At sufficiently cold gas temperatures, species can freeze out with a rate constant of
\begin{equation}
   k_{\mathrm{freeze},i} = S_i\left(1 + \beta_i \frac{1.671\times10^{-3}}{a_{\mathrm{grain}} T_{\mathrm{gas}} }\right)\sigma_\mathrm{grain}\sqrt{\frac{8k_\mathrm{B}T_\mathrm{gas}}{\pi m_i}},
\end{equation}
as described in \citet{rawlingsDirectDiagnosisInfall1992}, where $a_\mathrm{grain}$ is the grain radius (in cm) and $m_i$ is the mass of the species $i$ (in g), and the constant $1.671\times10^{-3}$ is in K$\,$cm. $\beta_i$ is 1 for positively charged species, and 0 for others, which captures the attractive interaction between negatively charged grains and positive ions, $\sigma_\mathrm{grain}$ is the cross section area of the grain per hydrogen nucleus, and is averaged between that of silicate and graphite: $\sigma_\mathrm{grain} = (7.908\times10^{-22}+8.473\times10^{-22})/2\;\mathrm{cm}^{2}$ \citep{cazauxMolecularHydrogenFormation2002, cazauxH2FormationGrain2004}. The sticking coefficient $S_i$ is assumed to be 1 for all species except those of atomic and molecular hydrogen, which are calculated according to \cite{chaabouniStickingCoefficientHydrogen2012}. The
fractional abundances of the species on the ices as they 
accumulate are shown in \Cref{fig:monolayers} for an isothermal 
cloud with $n_\mathrm{H,nuclei}=10^4\;\mathrm{cm}^{-3}$ and
$T=10\;\mathrm{K}$.

Once species are present on the surface of the ice, they can desorb back into
the gas phase via several mechanisms, broadly categorized into
non-thermal and thermal effects. 
Species can be desorbed via direct cosmic ray heating \citep{robertsDesorptionInterstellarIces2007}:
\begin{equation}
k_{\mathrm{cr}} = 4\pi\zeta \cdot \langle \pi a_\mathrm{grain}^2 n_\mathrm{grain} \rangle,
\end{equation}
where $\zeta$ is the CRIR in s$^{-1}$ (as above) and $n_\mathrm{grain}$ is the dimensionless abundance of dust grains relative to the total H-nucleus density, so that the bracketed term is a cross section per H nucleus (cm$^2$), consistent with $\sigma_\mathrm{grain}$ above.
Species can also interact with photons that are induced via cosmic rays or
directly radiated via the (interstellar) radiation field, resulting in two desorption terms:
\begin{equation}
k_{\mathrm{(cr)pd}} = \langle \pi a_\mathrm{grain}^2 n_\mathrm{grain} \rangle \cdot Y \cdot F_\mathrm{P} \cdot
\left(\zeta + \frac{F_\mathrm{UV}}{\eta} e^{-1.8 A_{\rm V}} \right),
\end{equation}
where $Y=0.03$ is the (dimensionless) yield per photon \citep{obergFormationRatesComplex2009}, the photon flux $F_\mathrm{P}=4875\;\mathrm{cm}^{-2}\,\mathrm{s}^{-1}$ \citep{cecchi-pestelliniCosmicRayInduced1992}, and the (dimensionless) attenuation factor for photons on the dust set to $\eta=10^3$. Both mechanisms only operate
if the binding energy limit is not exceeded, being $E_\mathrm{cr}=1210\;\mathrm{K}$ and $E_\mathrm{(cr)pd}=10^4\;\mathrm{K}$ for the reactions respectively.

Additionally, species can thermally desorb back to the gas-phase according to an Arrhenius process,
\begin{equation}
    k_{\mathrm{des},i} = \nu_{\mathrm{des},i}\exp\left(-E_{\mathrm{bind},i}/k_\mathrm{B}T_{\mathrm{dust}}\right),
\end{equation}
where $\nu_{\mathrm{des},i}$ is the desorption prefactor and $E_{\mathrm{bind},i}$ is in K. This is usually calculated as in \citet{hasegawaModelsGasGrainChemistry1992},
\begin{equation}
    \nu_{\mathrm{des},i}^{\mathrm{HH}} = \sqrt{\frac{2n_{\mathrm{sites}}k_{\mathrm{B}}E_{\text{bind},i}}{\pi^2m_i}},
\end{equation}
where $n_{\mathrm{sites}}$ (in cm$^{-2}$) is the surface site density on the grains. This neglects the contribution of the rotational partition function, which makes the prefactor temperature-dependent and results in an underestimation of $\nu_{\mathrm{des}}$ by multiple orders of magnitude for larger species \citep{minissaleDustInterstellarCatalyst2016, ligterinkOverviewDesorptionParameters2023}. 
Optionally, but not enabled by default, the desorption prefactor can be calculated using Transition State Theory \citep[TST, see][chap.~18]{atkinsAtkinsPhysicalChemistry2022}, according to \citet{ligterinkOverviewDesorptionParameters2023} and \citet{minissaleThermalDesorptionInterstellar2022}, which in turn are based on \citet{taitNalkanesPt111C0001Pt1112006},
\begin{equation}
    \nu_{\text{des},i}^{\text{TST}}=\frac{k_{\mathrm{B}}T_{\text{dust}}}{h}\frac{q^{\ddagger}_i}{q_i}=\frac{k_{\mathrm{B}}T_{\text{dust}}}{h}q_{\text{rot},i}^{\ddagger}q_{\text{tr,2D},i}^{\ddagger},
\end{equation}
where $q^{\ddagger}_i$ indicates the partition function of the transition state (gas-phase if desorption is barrierless), and $q_i$ indicates the partition function of the adsorbed species. Assumptions behind the second equality are discussed in \citet{minissaleThermalDesorptionInterstellar2022}. $q_{\text{tr,2D},i}^{\ddagger}$ is the 2-dimensional translational partition function, so only includes movement of the gas-phase species parallel to the surface. This can be calculated as 
\begin{equation}
    q_{\text{tr,2D},i}^{\ddagger} = \frac{A}{\Lambda_i^2}\text{, with }\Lambda_i = \frac{h}{\sqrt{2\pi m_i k_{\mathrm{B}}T_{\text{dust}}}},
\end{equation}
where $A$ is the area taken up by the species (in cm$^2$), assumed to be $n_{\mathrm{sites}}^{-1}$.
Here, it is assumed that the molecule has not yet thermalized to the gas temperature, such that its temperature is still $T_{\text{dust}}$. 

The rotational partition function of a non-linear molecule (in the high temperature approximation) can be calculated as 
\begin{equation}
    q_{\text{rot,3D},i}^{\ddagger} = \frac{\sqrt{\pi}}{h^3}\left(8\pi^2k_{\mathrm{B}}T_{\text{dust}}\right)^{3/2}\frac{\sqrt{I_{x,i}I_{y,i}I_{z,i}}}{\sigma_i},
    \label{eq:qrot3d}
\end{equation}
where $I_{x,i}$, $I_{y,i}$ and $I_{z,i}$ (in g$\,$cm$^2$) are the three principal moments of inertia for rotation of the species, and $\sigma_i$ is its symmetry factor, which is the number of indistinguishable rotations of the molecule. The moments of inertia can either be taken from a database, such as the Cologne Database for Molecular Spectroscopy \citep[CDMS,][]{mullerCologneDatabaseMolecular2001, mullerCologneDatabaseMolecular2005} or the Computational Chemistry Comparison and Benchmark DataBase \citep[CCCBDB,][]{johnsonComputationalChemistryComparison2002}, or can be calculated from the molecular geometry. The symmetry factor of a molecule $\sigma_i$ can be deduced from its point group \citep[see Table~140 of][]{herzbergInfraredRamanSpectra1987}.

For diatomic and linear molecules, which have two degenerate rotational modes, the rotational partition function can be calculated as 
\begin{equation}
    q_{\text{rot,2D},i}^{\ddagger} = \frac{1}{h^2}\left(8\pi^2k_{\mathrm{B}}T_{\text{dust}}\right)\frac{\sqrt{I_{y,i}I_{z,i}}}{\sigma_i}\text{, where }I_{y,i}=I_{z,i}.
    \label{eq:qrot2d}
\end{equation}
We note that Eq.~20 in \citet{minissaleThermalDesorptionInterstellar2022} contains an extra factor of $\sqrt{\pi}$, and values reported for $q_{\text{rot,2D}}^{\ddagger}$ in Table~4 are off by a factor of $\pi$.

At first, the species freeze out onto the bare grain surface, but as soon as the surface is covered in more than one monolayer of ice, the ice starts to grow radially, introducing a bulk volume. Abundances must then be corrected for the fact that material is stored in the bulk rather than at the surface \citep{garrodFormationCO2Other2011},
\begin{align}
\frac{\mathrm{d}n_{\mathrm{s}}(i)}{\mathrm{d}t}_{\mathrm{s}\rightarrow \mathrm{b}} &= -\min\left[1, \theta_{\mathrm s}\right] \frac{n_{\mathrm{s}}(i)}{n_\mathrm{s,tot}} \frac{\mathrm{d} n_\mathrm{s,tot}}{\mathrm{d}t}
\text{ if }\frac{\mathrm{d}n_\mathrm{s,tot}}{\mathrm{d}t} > 0,
\end{align}
where $\theta_\mathrm{s}$ is the (dimensionless) surface coverage fraction and $n_\mathrm{s,tot}$ is the total abundance of surface species with respect to hydrogen nuclei. So, upon net adsorption (i.e.~freeze-out is faster than desorption) some of the surface material is displaced to the bulk, depending on how much of the grain is already covered by ice. In the case of net desorption, some material in the bulk has to be moved to the surface,
\begin{align}
\frac{\mathrm{d}n_{\mathrm{b}}(i)}{\mathrm{d}t}_{\mathrm{b}\rightarrow \mathrm{s}} &= \hphantom{-}\min\left[1, \frac{n_\mathrm{b,tot}}{n_\mathrm{s,tot}}\right] \frac{n_{\mathrm{b}}(i)}{n_\mathrm{b,tot}} \frac{\mathrm{d} n_\mathrm{s,tot}}{\mathrm{d}t}\text{ if }\frac{\mathrm{d}n_\mathrm{s,tot}}{\mathrm{d}t} < 0,
\end{align}
where $n_\mathrm{b,tot}$ is the total bulk abundance with respect to hydrogen nuclei. The slightly different (in the case of net desorption) formulation of \citet{hasegawaThreephaseChemicalModels1993} can also optionally be used. We then also include random swapping between the bulk back to the surface to account for the fact that species can move around due to thermal effects: $k_{\mathrm{swap},i}=v_{\mathrm{diff},i}\exp(-E_{\mathrm{bind},i}/T_\mathrm{dust})$. 

At low temperatures and high densities, astrochemical models overpredict the abundance of \ch{H2} ice because of its high abundance in the gas-phase, and so high freeze-out rate. The binding energy of \ch{H2} on an \ch{H2} surface is much lower than that of \ch{H2} on \ch{H2O} \citep{cuppenSimulationFormationMorphology2007, dasEffectBindingEnergies2021}, which means that no multilayer ice of \ch{H2} would really form. To remedy this, we use the encounter desorption mechanism \citep{hincelinNewSimpleApproach2015}, which causes \ch{H2} to more easily desorb when it diffuses over another \ch{H2} molecule. This requires the \ch{H2} molecule to go from a \ch{H2O} surface, where it is relatively strongly bound, to a \ch{H2} surface, where it is very weakly bound. This means that this process is endothermic and would happen slowly, so should not be considered a real physical process but a way to solve the \ch{H2} ice overabundance.

\subsubsection{Reactions on the grains}\label{sec:grainchem}
\begin{figure}
    \centering
    \includegraphics[width=1.0\linewidth]{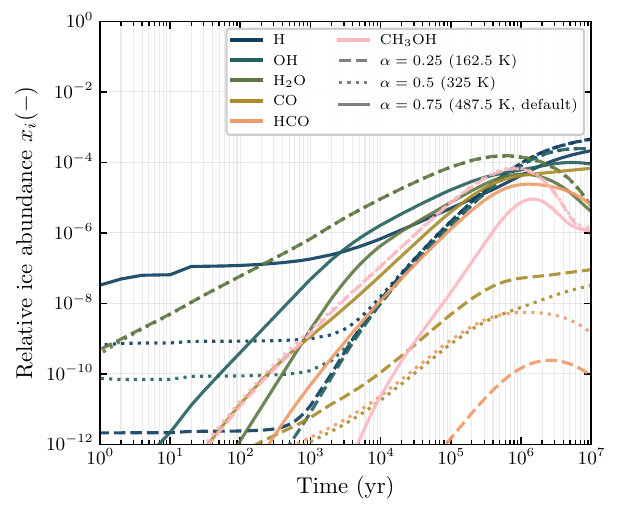}
    \caption{The formation of ice species from \Cref{fig:monolayers} with the hydrogen diffusion barrier $E_{\text{diff,H}}=\alpha E_{\text{bind,H}}$, for $\alpha=0.25$, 0.5, and 0.75.}
    \label{fig:iceformation}
\end{figure}

The diffusion (hopping) rate constant is calculated as 
\begin{equation}
    k_{\mathrm{diff},i} = \nu_{\mathrm{diff},i} \exp\left(-E_{\mathrm{diff},i}/k_\mathrm{B}T_{\mathrm{dust}}\right),
\end{equation}
where $E_{\mathrm{diff},i}$ is the diffusion barrier of species $i$, and $\nu_{\mathrm{diff},i}$ is the diffusion prefactor. The diffusion barrier is often assumed to be a fixed fraction of the binding energy. However, this fraction is not consistent among species \citep{furuyaDiffusionActivationEnergy2022, ligterinkMolecularMobilityExtraterrestrial2025, benedettiCODiffusionInterstellar2025}. Assumptions about diffusion barriers have a large effect on the predicted abundances \citep{dijkhuisSensitivityAnalysisInterstellar}, so we have decoupled the binding energies and diffusion barriers. The effect of changing the hydrogen diffusion barrier on the abundances of some ice species can be seen in \Cref{fig:iceformation}. Observing that only the highest diffusion
to binding ratio is consistent with producing more \ch{CO} than \ch{CH3OH},
it is thus chosen as the default. 

The most important type of reaction of species on the grain is the Langmuir-Hinshelwood (LH) reaction. Two species diffusing on a grain meet each other and can react. The rate for LH reactions is
\begin{equation}
\frac{\mathrm{d} n_s(i)}{\mathrm{d}t} = - \frac{f_{\mathrm{comp}}\left(k_{\mathrm{diff},i}+k_{\mathrm{diff},j}\right)} {N_\mathrm{sites} n_\mathrm{d}} n_{\mathrm{s}}(i) n_{\mathrm{s}}(j),
\end{equation}
where $N_{\mathrm{sites}}$ is the (dimensionless) number of adsorption sites per dust grain and $n_\mathrm{d}$ is the dust grain number density (cm$^{-3}$). $f_{\mathrm{comp}}$ captures the competition between the reaction, and either of the reactants diffusing or desorbing away \citep{changGasgrainChemistryCold2007},
\begin{equation}
    f_{\mathrm{comp}} = \frac{\kappa_{ij}}{\kappa_{ij}+k_{\mathrm{diff},i}+k_{\mathrm{diff},j}+k_{\mathrm{des},i}+k_{\mathrm{des},j}},
\end{equation}
\begin{equation}
    \kappa_{ij} = \max\left[\nu_{\mathrm{diff},i},\nu_{\mathrm{diff},j}\right]P_{\mathrm{reac},ij},
\end{equation}
where $P_{\mathrm{reac},ij}$ is the probability for crossing the barrier, either classically or through tunnelling,
\begin{equation}
        P_{\text{reac},ij} = \max\left[\exp\left(-\frac{E_{\text{reac},ij}}{k_\mathrm{B}T_{\text{dust}}}\right),\exp\left(-\frac{2a}{\hbar}\sqrt{2\mu_{ij} k_{\mathrm{B}}E_{\text{reac},ij}}\right)\right],
\end{equation}
where the tunnelling barrier width $a=1.40\times10^{-8}\;\mathrm{cm}$ (fixed) and the reduced mass $\mu_{ij}$ is in g.

Eley-Rideal (ER) reactions can occur when species $i$ on the ice surface is hit by a species $j$ from the gas-phase,
\begin{equation}
    \frac{\mathrm{d} n_{\mathrm{s}}(i)}{\mathrm{d}t} = - k_{\mathrm{freeze},j}n_j P_{\mathrm{reac},ij}\frac{n_{\mathrm{s}}(i)}{n_{\mathrm{s,tot}}},
\end{equation}
where the division by the total surface abundance is offset by a small epsilon (set at $10^{-30}$ by default) to prevent division by zero.
These reactions can release excess energy to the products, which can cause them to desorb. This process is called chemical (also referred to as reactive) desorption. In \texttt{UCLCHEM} 4.0, we use a combination of \citet{minissaleDustInterstellarCatalyst2016} for chemical desorption from bare grains, and \citet{fredonQuantificationRoleChemical2021} for chemical desorption from ices. See \citet{dijkhuisSensitivityAnalysisInterstellar} for details.

Additionally, \texttt{UCLCHEM} includes the photon and cosmic ray induced reactions
on the surface and in the bulk. The reactions are copied from the gas-phase (i.e.~from UMIST or KIDA), with
the phase of the reactants and products changed to surface and bulk species respectively.
The rates of the reaction are modified to account for the fact that ice species have a lower dissociation cross-section \citep[with a factor of 0.3 by default, as found by][]{kalvansEfficiencyPhotodissociationMolecules2018}, and species in the bulk are shielded by species above it \citep{kalvansCosmicRayInducedDiffusion2014}.

In order to model the formation of molecular hydrogen on the grains, two updated
models are available in the framework. The first treatment is based on 
\citet{cazauxMolecularHydrogenFormation2002, cazauxH2FormationGrain2004}
adapted from the Photon-dominated Regions (PDR) chemical code \texttt{UCLPDR} \citep{bellChemistryTransientMicrostructure2005, bellMolecularLineIntensities2006, bayetChemistryCosmicRay2011, priestleyModellingArHEmission2017}. It computes the thermal velocity, sticking
factor, cross section and formation efficiency for both silicate
and graphite grains, their product provides the $\ch{H2}$ formation rate. 
The exact treatment can be found in \Cref{app:h2form}. Alternatively, we 
use the formation of molecular hydrogen via direct reactions on the grain,
using the desorption mechanisms discussed in \Cref{sec:surfbulkinteraction} to model the feedback of \ch{H2} into the gas phase. We adopt the
direct computation of the \ch{H2} formation as the default method at cold
temperatures ($T_\mathrm{dust}\leq150\;\mathrm{K}$) as $R_{\ch{H2},\mathrm{LH,ER}}=R_{\ch{H2},\mathrm{LH}}+R_{\ch{H2},\mathrm{ER}}$ and the Cazaux \& Tielens (CT) approach 
above said threshold as $R_{\ch{H2},\mathrm{CT}}$. The user is free to either modify that threshold or choose
only one of the mechanisms.

\section{The physics}
\label{sec:phys}
In order to model the chemistry of interstellar environments, the physics has to be accounted for. Since most rate constants $k$ depend on the physical conditions ($k(T,F_\mathrm{UV},\zeta,...)$), it is important that the physics accurately models the environment. Two common approaches can be taken to simulate the physics of these environments: dynamical simulations and parametrized profiles. 
The former tries to directly capture the effect of gravity, (magneto)hydrodynamics, radiation, and stellar evolution, sometimes even coupling the effect of chemical reactions, heating and cooling directly to the dynamical simulation. 
This coupling introduces a large computational overhead, as for each particle or cell in the simulation the chemistry has to be evaluated. 
The latter method instead focuses on trying to find parametrized models of these complex environments, reducing the
problem to a lower dimension.

\subsection{Physical modelling of astrophysical environments}
\begin{figure}
    \centering
    \includegraphics[width=1\linewidth]{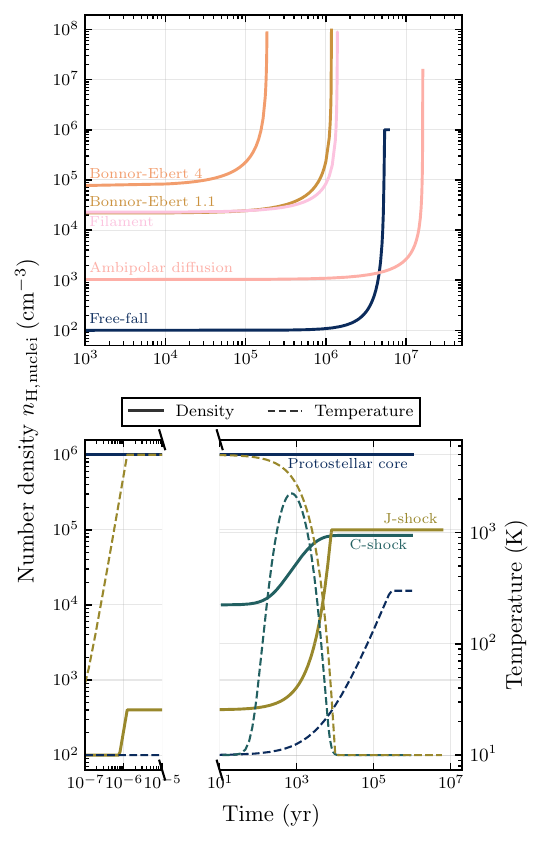}
    \caption{Illustration of the different physical parametrizations available in \texttt{UCLCHEM}. The upper plot
    shows the time evolution of the density in the different stage 1 collapse models. The bottom plot 
    shows the corresponding temperature and density profiles of the stage 2 models.}
    \label{fig:modelstages}
\end{figure}

\begin{figure}
    \centering
    \includegraphics[width=1\linewidth]{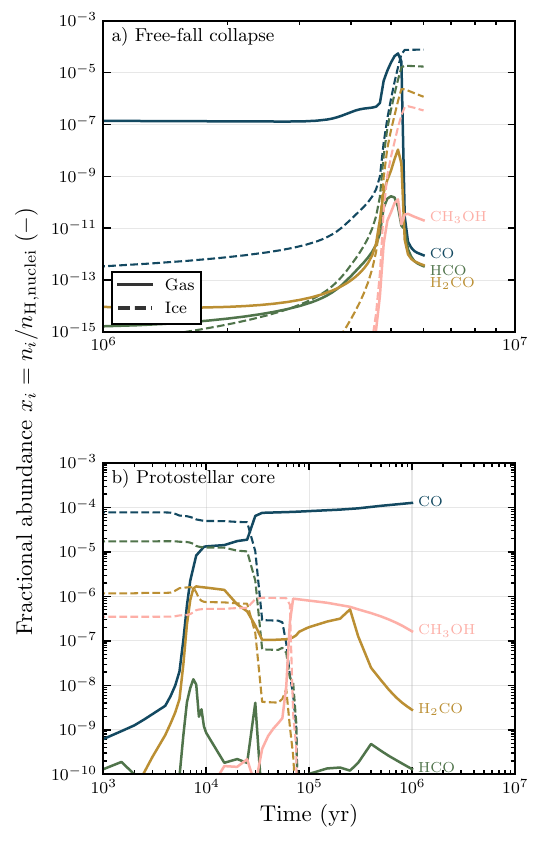}
    \caption{Example of a typical \texttt{UCLCHEM} modelling setup, a free-fall collapse stage that takes 5 Myr and a
    protostellar core warm-up stage that reaches $T_{final}=300\;\mathrm{K}$ in $2\times10^5$years.}
    \label{fig:a_chemistry}
    
\end{figure}

In order to accurately model the evolution of astrophysical environments,
\texttt{UCLCHEM} provides distinct parametrizations of density, temperature and supplementary 
physical treatment over time. \texttt{UCLCHEM} models typically consist of two stages, but the
number of models run consecutively can be increased in order to model more complex scenarios.
The first stage is the collapse stage (\Cref{sec:collapse_models}), where the model evolves from a diffuse cloud with 
only atomic initial abundances into a dense molecular cloud that has formed molecules in the gas-phase, and,
given a low enough temperature, ices as well. The second stage then tries to
model the effect of internal or external sources of energy being introduced. This can be
in the form of the formation of a protostellar core (\Cref{sec:protostellar_models}) that radiates energy, a slow
C-type shock (\Cref{cshock_models}) or a faster J-type shock (\Cref{jshock_models}). A brief illustration of the different physical parametrizations
can be found in \Cref{fig:modelstages} and the corresponding chemistry of a typical free-fall collapse with
subsequent protostellar core is shown in \Cref{fig:a_chemistry}.


\subsubsection{Collapse models}
\label{sec:collapse_models}
In order to simulate the collapse of diffuse material to a denser molecular phase, 
\texttt{UCLCHEM} includes several collapse models, from  free-fall
collapse  that can be used with arbitrary initial and final densities, to less 
flexible collapse models that are calibrated using hydrodynamical simulations.
The default collapse models a free-fall collapse that evolves the density with time according to
\begin{equation} \label{eq:ngas_collapse}
\frac{\mathrm{d} n}{\mathrm{d} t}=B\left(\frac{n^4}{n_0}\right)^{1 / 3}\left\{24 \pi G m_{\mathrm{H}} n_0\left[\left(\frac{n}{n_0}\right)^{1 / 3}-1\right]\right\}^{1 / 2},
\end{equation}
where $n_0$ is the initial density and $B$ takes values between 0 and 1 \citep{rawlingsDirectDiagnosisInfall1992}. $B = 1$ corresponds to pure free-fall, with $B < 1$ approximating the effects of pressure and/or magnetic fields slowing the collapse.

Less-idealised collapse models are implemented using analytical approximations to the results of (magneto)hydrodynamic (MHD) simulations developed by \citet{priestleyEfficientMethodDetermining2018}. 

In these collapse models, the density is given as a function of radius $r$ and time $t$ by
\begin{equation}
    n(r,t) = \frac{n_0(t)}{1 + [r/r_0(t)]^{a(t)}},
    \label{eq:profile}
\end{equation}
where the functions $n_0(t)$, $r_0(t)$ and $a(t)$ have been chosen to reproduce the density profile evolution of simulated collapsing prestellar cores to within $10\%$. Four collapse models are available: the one-dimensional hydrodynamic collapse of a Bonnor-Ebert sphere with a moderate ($1.1$) or major ($4$) density enhancement, as studied by \citet{aikawaMolecularEvolutionCollapsing2005}; the MHD collapse of a fragment of a magnetised filament, from \citet{nakamuraFragmentationFilamentaryMolecular1995}; and the non-ideal MHD collapse via ambipolar diffusion of a uniform-density core, initially in a magnetically-stable regime, from \citet{fiedlerAmbipolarDiffusionStar1993}. For the latter two cases, the model provides the density profile perpendicular to the initial magnetic field direction (i.e.~in the midplane of the resulting disc or pseudodisc).

When using these collapse models, it is necessary to specify an initial grid of radial points, for which the  density is calculated at $t=0$. At each subsequent timestep, a new radius is calculated for each point, corresponding to the radial distance moved in that interval. The density of the point is then updated according to \Cref{eq:profile}. For the two Bonnor-Ebert collapse models, the radius update is done by mass conservation (the mass enclosed by the point should not vary with time). For the two MHD collapses, fits to the radial velocity profiles, $v_r(r,t)$, are obtained in a similar manner to \Cref{eq:profile}, and used to determine the distance moved by each point.

It is important to note that all four simulations assume an isothermal equation of state, meaning that the collapses end in a singularity at the centre at some finite time. This is represented by the expressions for $n_0(t)$ and $r_0(t)$ going to infinity and zero, respectively, at a critical value of $t$ which depends on the model. Attempting to run a collapse model beyond this effective final time will result in unphysical behaviour, so \texttt{UCLCHEM} effectively stops the model at a maximum density of $n_\mathrm{H,nuclei}=10^{8}$~cm$^{-3}$ at the final time. It is then possible to evolve the chemistry
beyond this time at a constant density, but this is disabled by default.

\subsubsection{Protostellar warm-up models}
\label{sec:protostellar_models}
\begin{figure}
    \centering
    \includegraphics[width=1.0\linewidth]{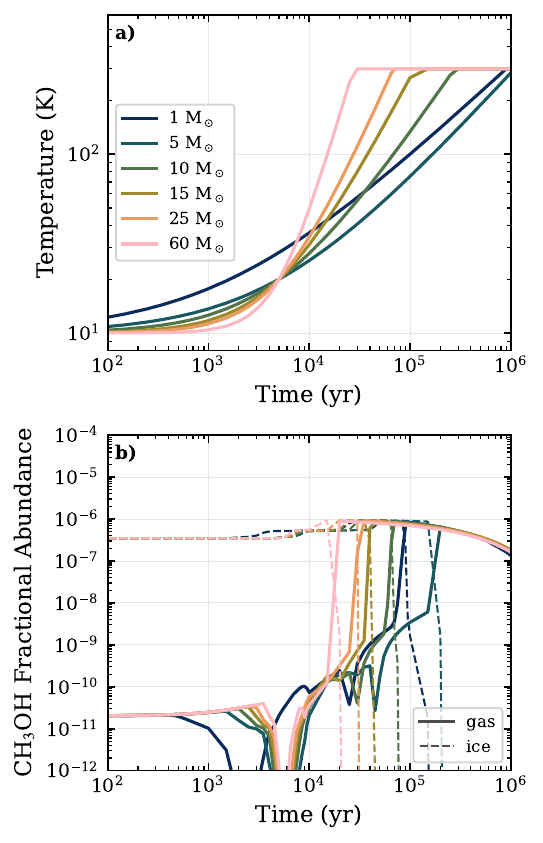}
    \caption{The different timescales of the zero-dimensional protostellar models with a maximum temperature
    of $T=300\;\mathrm{K}$. a) Shows the temperatures increasing over time with the different parametrizations and b) shows the gas and ice fractional
    abundances of methanol as the protostellar model evolves.}
    \label{fig:zerodtempprofiles}
\end{figure}

In the zero-dimensional (0D) protostellar module the key parameter is the temperature of the dust (and gas, if coupled), which varies with time as a function of the final mass of the protostar that is being modelled. Chemically this implies that different species sublimate at different times. In order to simulate the heating as a function of time and stellar mass, we assume the presence of an infrared source 
(usually assumed to be in the centre of the core) and this source will yield an increase in gas and dust temperature. The latter is a function of the source luminosity (and hence age) which in turn is a function of the mass. The approach we use is that described in \citet{vitiEvaporationIcesMassive2004} and \citet{awadWarmCoresRegions2010}, namely:
\begin{equation} \label{eq:Td_hotcore}
    T(t,d)= 10 + A t^B\sqrt{\frac{R}{d}}\text{~K},
\end{equation}
where  $T(t,d)$ is the temperature profile of gas and dust in the
gas surrounding the protostar, $t$ is the  age of
the  gas (in yr), $d$ is the distance from the core centre and $R$ is
the core radius. $A$ and $B$ are two constants derived from the boundary
conditions for different masses, see \Cref{tab:protostellarab}, derived from \citet{vitiEvaporationIcesMassive2004} and \citet{awadWarmCoresRegions2010} for
the zero-dimensional case and refitted in \citet{tramOnedimensionalTimedependentModelling2026} for the one-dimensional case.
These temperatures are then limited to a user-defined maximum for the 
zero-dimensional protostellar model, set at $T_{\rm max}=300\;\mathrm{K}$ by default. The effective temperature profiles
for the zero-dimensional model and its effect on
the timescale of methanol desorbing into the gas-phase can be found in \Cref{fig:zerodtempprofiles}.

\begin{table}
\centering
\caption{Temperature profile parameters for protostellar models \citep{vitiEvaporationIcesMassive2004, awadWarmCoresRegions2010,tramOnedimensionalTimedependentModelling2026}}
\label{tab:protostellarab}
\begin{tabular}{|c|c|c|c|c|}
\hline
 \multirow{2}{*}{\begin{tabular}{c}Mass \\ (M$_\odot$)\end{tabular}} & 
\multicolumn{2}{c|}{Zero-dimensional} & \multicolumn{2}{c|}{One-dimensional} \\
\cline{2-5}
& $A\;(\mathrm{K})$ & $B\;(-)$ & $A\;(\mathrm{K})$ & $B\;(-)$ \\
\hline
1 & $1.927 \times 10^{-1}$ & 0.5339 & $3.1417 \times 10^{-2}$ & 0.5329 \\
\hline
5 & $4.856 \times 10^{-2}$ & 0.6255 & $3.5495 \times 10^{-2}$ & 0.5324 \\
\hline
10 & $7.847 \times 10^{-3}$ & 0.8395 & $4.9653 \times 10^{-4}$ & 0.9 \\
\hline
15 & $9.697 \times 10^{-4}$ & 1.085 & $9.5928 \times 10^{-4}$ & 0.9 \\
\hline
25 & $1.706 \times 10^{-4}$ & 1.289 & $1.4158 \times 10^{-3}$ & 0.9 \\
\hline
60 & $4.74 \times 10^{-7}$ & 1.98 & $2.817 \times 10^{-3}$ & 0.9 \\
\hline
\end{tabular}
\end{table}

We refer the reader to \Cref{sec:1drad} for the full one-dimensional treatment that expands
the protostellar models with a more accurate temperature profile based on
the internal radiation source.

\subsubsection{C-shock models}
\label{cshock_models}
Continuous (C-type) shocks occur in strongly magnetized media where ion-neutral decoupling causes the physical properties of the gas, including velocity, density, and temperature, to change gradually across the shock. The C-type shock implementation in \texttt{UCLCHEM} (\Cref{fig:modelstages}) follows the parametrization described by \citet{jimenez-serraParametrizationCshocksEvolution2008}, which describes the evolution of the ion and neutral fluid velocities as a function of time and derives the physical structure of the shock through analytic prescriptions calibrated against detailed MHD shock models calculated with \texttt{mhd\_vode} \citep{flowerInfluenceGrainsPropagation2003}.

Since \texttt{UCLCHEM} employs a parametrized description of the shock structure, the shock type is not determined self-consistently by the code \citep{vanlooTimedependentSimulationsSteady2009, grassiChallengesModellingMicrophysics2019}. Instead, the nature of the shock must be specified through the physical parameters supplied by the user. In particular, the magnetic field strength enters the parametrization through the parameter \texttt{bm0}, which represents the magnetic field in units of $\muup$G following the scaling relations adopted in \citet{draineMagnetohydrodynamicShockWaves1983}. Appropriate values of this parameter must therefore be chosen to ensure that the physical conditions correspond to those required for a C-type shock.

The shock develops across a dissipation region over which the physical properties of the gas evolve smoothly. In the current implementation the spatial extent of this region scales with the shock velocity and the pre-shock density following this prescription. The corresponding crossing time, $t_\mathrm{diss}$, defines the duration of the shock stage, after which the gas is considered to have entered the post-shock regime. Once this stage is reached, the gas and dust temperatures are not allowed to fall below a user-defined parameter, \texttt{minimum\_temperature}. By default this value corresponds to the initial temperature of the medium, but it can be increased to mimic additional heating of the post-shock gas. A schematic of the shock and post-shock stages used in the model is shown below:
\begin{center}
\begin{tabular}{c c}
$t \le t_\mathrm{diss}$ & $t > t_\mathrm{diss}$ \\
\hline
shock & post-shock
\end{tabular}
\end{center}

The velocity evolution yields the position within the shock, the ion and neutral fluid velocities, and the resulting ion-neutral drift velocity. The neutral gas temperature is described by
\begin{equation}
T_\mathrm{n} = T_\mathrm{init} +
\frac{(a_T z_\mathrm{n})^{b_T}}
{\exp(z_\mathrm{n}/z_3)-1}\text{ K},
\end{equation}
where $z_\mathrm{n}$ (in cm) represents the position along the propagation of the shock. In the current implementation $b_T$ is fixed to 6, $z_3$ (in cm) is set to one sixth of the distance at which the neutral temperature reaches its maximum, and the normalization constant $a_T$ is derived so that the parametrized profile reaches the required maximum temperature $T_\mathrm{max}$. The neutral temperature defines the gas temperature used by the chemical network, and the dust temperature is assumed to be equal to the gas temperature.

The ion temperature includes an additional contribution from ion-neutral drift heating and is given by
\begin{equation}
T_\mathrm{i} = T_\mathrm{n} + \frac{m v_\mathrm{d}^{2}}{3k_{\rm B}}\text{ K},
\end{equation}
where $v_\mathrm{d}$ is the ion-neutral drift velocity (in km~s$^{-1}$, converted internally to cm~s$^{-1}$) and $m$ is the mean particle mass (in g; taken as twice the mass of a hydrogen nucleus for the neutral fluid).

The neutral density is given by
\begin{equation}
n_\mathrm{n} = \frac{n_0 v_\mathrm{s}}{v_\mathrm{s}-v_\mathrm{n}} \text{ cm}^{-3},
\end{equation}
where $n_0$ is the pre-shock density (cm$^{-3}$), and $v_\mathrm{s}$, the shock velocity, and $v_\mathrm{n}$, the neutral fluid velocity, are both in the same velocity units, and are typically expressed in km~s$^{-1}$. This relation describes the gradual compression of the gas as the neutral fluid decelerates through the shock.

Sputtering of icy grain mantles is included during the shock stage. In the \texttt{UCLCHEM} implementation the sputtering rates follow the formulation described in Appendix~B of \citet{jimenez-serraParametrizationCshocksEvolution2008}. The rate of mantle removal is computed for a set of impacting species and integrated over the timestep of the model. The resulting mantle loss is distributed among the surface species in proportion to their abundances on the grain. In the current implementation sputtering is restricted to volatile mantle species unless the relative velocity exceeds 19~km~s$^{-1}$, above which refractory species are also released following \citet{guilletShocksDenseClouds2011}. The same sputtering routine is used in both shock modules, with the relative velocity supplied by the respective shock parametrization.

\subsubsection{J-shock models}
\label{jshock_models}
Jump (J-type) shocks are characterized by an abrupt discontinuity in the physical properties of the gas, resulting in rapid increases in temperature and density at the shock front. The J-shock module in \texttt{UCLCHEM} (\Cref{fig:modelstages}) builds upon the methodology used for the successful C-shock implementation, adopting a parametrized description of the shock evolution. The parametrization introduced by \citet{jamesTracingShockType2020} was calibrated using a grid of shock models calculated with \texttt{mhd\_vode} \citep{flowerInterpretingObservationsMolecular2015} and describes two stages of shock evolution: the shock-front phase and the subsequent post-shock relaxation layer. The structure of the J-shock module is illustrated below:

\begin{center}
\begin{tabular}{c c c}
$t=0$ & $t=t_\mathrm{shock}$ & $t=t_\mathrm{cool}$ \\
\hline
shock front & cooling phase & post-shock gas
\end{tabular}
\end{center}

The duration of the shock front is determined from a characteristic mean free path, and in units of s, is described as follows:
\begin{equation}
t_\mathrm{shock} =
\frac{1}{\sqrt{2}\,\pi\,(2.4\times10^{-8}\,\mathrm{cm})^2\,10^{12}\,(v_\mathrm{s}/\mathrm{km\,s^{-1}})}\ \mathrm{s},
\end{equation}
where $v_\mathrm{s}$ is the initial velocity of the shock, taken in km~s$^{-1}$ in the equation above; the constant $10^{12}$ combines a fixed reference pre-shock density of $10^{3}\,\mathrm{cm^{-3}}$ with unit-conversion factors and is not scaled by the model's actual pre-shock density. The increase in temperature and density within the shock front, at times $t<t_{\rm shock}$ is described by the following:
\begin{equation}
\label{eq:temperature_jshock}
T = T_\mathrm{max}\left(\dfrac{t}{t_\mathrm{shock}} \right)^2,\
\end{equation}
\begin{equation}
\label{eq:density_jshock}
    n_\mathrm{H} = 4n_\mathrm{H,initial}\left(\dfrac{t}{t_\mathrm{shock}} \right)^3,\
\end{equation}
where $n_\mathrm{H,initial}$ is the initial pre-shock number density in cm$^{-3}$. An assumption is made that density, $n_\mathrm{H,nuclei}$, during the shock front increases to about four times its initial value, assuming the Rankine-Hugoniot relations, whilst the temperature $T$ increases to its maximum obtainable value, $T_\mathrm{max}$, which is described by $T_\mathrm{max} = 5\times10^3\left(\dfrac{v_\mathrm{s}}{10\;\mathrm{km\,s^{-1}}}\right)^2$~K \citep{williamsObservationalMolecularAstronomy2013}, with $v_\mathrm{s}$ again in km~s$^{-1}$.

After the shock-front phase, the shocked gas enters a post-shock relaxation layer in which it cools while continuing to compress. This phase lasts until the cooling timescale, $t_\mathrm{cool}$:
\begin{equation}
\label{eq:t_cool}
t_\mathrm{cool} = \dfrac{t_\mathrm{year}\times10^6}{n_\mathrm{H,initial}}\ \mathrm{s},
\end{equation}
where $t_\mathrm{year}$ is the number of seconds in one year and $n_\mathrm{H,initial}$ is the pre-shock number density in cm$^{-3}$. This relation represents the parametrized scaling derived from polynomial fits to shock timescales measured from the \texttt{mhd\_vode} models. The changes in the temperature and density in the relaxation layer follow:
\begin{equation}
T = T_\mathrm{max} \exp\left[-\lambda_\mathrm{T}\left(\dfrac{t}{t_\mathrm{cool}} \right)\right],\
\end{equation}
\begin{equation}
    n_\mathrm{H} = 4n_\mathrm{H_\mathrm{initial}}\exp\left[\lambda_\mathrm{n_\mathrm{H}}\left(\dfrac{t}{t_\mathrm{cool}} \right)\right].\
\end{equation}
The gas temperature therefore decreases exponentially. Meanwhile, the density increases towards \(n_\mathrm{H,max}\) derived from \texttt{mhd\_vode} models, as
\begin{equation}
n_\mathrm{H_\mathrm{max}} = \left(\left(v_\mathrm{s}/\mathrm{km\,s^{-1}}\right) \times n_\mathrm{H,initial} \right) \times 10^2\ \mathrm{cm^{-3}}.\
\end{equation}
The $\lambda$ constants are described as:
\begin{equation}
\lambda_\mathrm{T} = \mathrm{ln}\left(\dfrac{T_\mathrm{max}}{T_\mathrm{initial}} \right),\
\end{equation}
\begin{equation}
\lambda_\mathrm{n_\mathrm{H}} = \mathrm{ln}\left(\dfrac{n_\mathrm{H_\mathrm{max}}}{4 n_\mathrm{H_\mathrm{initial}}} \right).\
\end{equation}
In the very last part of the J-shock models, once the current time exceeds the defined $t_\mathrm{cool}$, an assumption is made that the gas has cooled back to its initial temperature $T_\mathrm{initial}$. 

\subsection{Cosmic ray treatment}
Cosmic rays (CRs) are very important contributors to the chemistry in the dense gas in the ISM, as they kick start chemistry at magnitudes higher than $\sim$ 3 where UV photons do not penetrate. While the chemical reactions due to interactions of CRs with atoms and molecules are included in all the chemical databases that are inputs to chemical models, the handling of CRs in \texttt{UCLCHEM} has only recently been improved by introducing the CR ionization rate and the H$_2$ dissociation rate as a function of column density following the recipe of \citet{padovaniCosmicrayIonisationCircumstellar2018}.  The details of the above treatment can be found in \citet{odonoghueEffectsCosmicRays2022} but here we report the equation that deals with the polynomial fit developed in \citet{padovaniCosmicrayIonisationCircumstellar2018} to determine the CR ionization rate as well as the H$_2$ dissociation rate as a function of the column density,
\begin{equation}
    \log_{10}\frac{\zeta}{s^{-1}}=\sum_{k=0}^9 c_k \log_{10}\left(\frac{N_\mathrm{H}}{\mathrm{cm}^{-2}}\right)^k,
\end{equation}
where $N_\mathrm H$ is the column density and $\zeta$ is the resulting CRIR used throughout the rest of this work (\Cref{sec:chem}). The fitting coefficients $c_k$ can be found in Table A1 of \citet{odonoghueEffectsCosmicRays2022}.

\subsection{Modeling radiation along the radius}
\label{sec:1drad}

\texttt{UCLCHEM} can also support one-dimensional modelling of clouds of gas. While the
zero-dimensional model simplifies the computation, it does come at the cost of not being able
to understand the layered nature of the astronomical  object. 
As the cumulative density of the gas along the path of the radiation increases, the radiation is increasingly attenuated, 
influencing the rates of the photochemistry.
By default, the visual extinction is computed based
on an external radiation field, taking the form:
\begin{equation}
A_{\mathrm{V},i} = \sum_{j\geq i} \frac{N_{\mathrm{H,nuclei},j}}{1.6\times 10^{21}}+ A_\mathrm{V,base}=\sum_{j\geq i} \frac{\Delta r_{\rm j} n_{{\rm H,nuclei},j}}{1.6\times 10^{21}}+ A_\mathrm{\rm V,base}\;,
\end{equation}
where $N_\mathrm{H,nuclei}$ is in cm$^{-2}$, the constant $1.6\times10^{21}$ is in cm$^{-2}\,\mathrm{mag}^{-1}$\citep{williamsObservationalMolecularAstronomy2013}, $j$ sums from the number of sampled points to
0, adding up the column densities from the outer edge
towards the point and $A_\mathrm{V,base}$ is the edge visual extinction, accounting for the extinction of the outer envelope. For zero-dimensional models, this simplifies to 
\begin{equation}
A_\mathrm{V} = \frac{r_{\rm out} n_{{\rm H,nuclei}}}{1.6\times 10^{21}}+ A_\mathrm{\rm V,base}.
\end{equation}

In \texttt{UCLCHEM} 4.0 we include an
 alternative one-dimensional model that accounts
 for both internal and external radiation for a typical
 collapse and protostellar model sequence \citep{tramOnedimensionalTimedependentModelling2026}.

The first stage starts with a collapse profile that
follows a free-fall collapse, with the visual extinction
computed using the aforementioned one-dimensional relationship. Each parcel in this model is evolved
until the final outermost particle reaches its
final density. 

The distance-dependent density at $r$ from the center uses the Bonnor-Ebert sphere with a maximum density:
\begin{equation}\label{eq:ngas}
    n_{\rm gas}(r)=\left\{
    \begin{array}{l l}
        n_{\rm 0} & \quad {\rm ~for~} r\le r_{\rm 0},\\
        n_{\rm 0}\left(\frac{r}{r_{\rm 0}}\right)^{-\alpha} & \quad {\rm ~for~} r>r_{\rm 0},
    \end{array}\right.
\end{equation}
where $n_{0}$ (in cm$^{-3}$) is the density in the center, $r_{0}$ (in cm) is the distance where $n_{0}$ remains constant, after which the density drops with a slope of $\alpha$. The evolution of $n_{\rm gas}$ from an initial density to this maximum value is governed by \Cref{eq:ngas_collapse}. 

Once the final parcel has reached its final
density, we proceed directly\footnote{Alternative stopping conditions are available, either stopping each parcel once
it reaches its individual maximum density or only stopping once a certain time beyond the collapse is reached, allowing the slowest collapse to evolve further.} to the protostellar
warm-up phase, accounting for both the internal and external
radiation. The expression for the extinction of the internal
radiation source is given by:
\begin{equation}
A_\mathrm{v}^{\text {int }}(r) = \frac{n_0}{1.6\times10^{21}}\begin{cases} r & \text { for } r \leq r_0, \\
r_0+\frac{r_0}{\alpha-1}\left[1-\left(\frac{r}{r_0}\right)^{-\alpha+1}\right] & \text { for } r>r_0,\end{cases} 
\end{equation}
and the extinction of the external radiation field is given by:
\begin{equation}
A_\mathrm{v}^{\text {ext }}(r) = \frac{n_0}{1.6\times10^{21}}\begin{cases} r_0\left(\frac{\alpha}{\alpha-1}-\frac{r}{r_0}\right) & \text { for } r<r_0, \\
\frac{r_0}{\alpha-1}\left(\frac{r}{r_0}\right)^{1-\alpha} & \text { for } r>r_0.\end{cases}
\end{equation}
with $n_0$ in cm$^{-3}$, $r$, $r_0$ in cm, and the constant $1.6\times10^{21}$ in cm$^{-2}\,\mathrm{mag}^{-1}$ as above.

The intensity of the external interstellar radiation field (ISRF) at the outer radius is characterised by a parameter $F_\mathrm{UV,0}$ (in Habing unit). At a given distance $r$ or gas density, the radiation field is attenuated as
\begin{equation}
    F_{\mathrm{UV},\text{\{int,ext\}}} = F_{\mathrm{UV},0,\text{\{int,ext\}}}\exp\left({-1.8A_{\rm V}^\text{\{int,ext\}}}\right),
\end{equation}
accounting for both the attenuation of the internal radiation of the protostellar stage as well as the external 
radiation in both stages.

For the protostellar stage, the internal radiation field is determined by the radiation of a luminous central source as
\begin{equation} \label{eq:Ustar}
        U(T_{\ast}) = \frac{\int^{20\,\rm \mu m}_{0.091\,\rm \mu m} u_{\lambda}(T_{\ast})e^{-\tau_{\lambda}}d\lambda}{u_{\rm ISRF}},
\end{equation}
where $u_{\lambda}(T_{\ast})=L_{\lambda}(T_{\ast})/(4\pi r^{2}c)$ is the spectral energy density of the central source with a luminosity $L_{\ast}$, $u_{\rm ISRF}=8.64\times 10^{-13}$~erg~cm$^{-3}$ is the energy density of the interstellar radiation field, and $\tau$ is the optical depth. We note that in proximity to the central luminous source, dust particles are destroyed (no dust survives) if the temperature is higher than the sublimation temperature of the dust grain (e.g.~$T_{\rm threshold} \sim 1500$~K for silicate grains; \citealt{hoangOriginDestructionRelativistic2015}). This distance, $r_{\rm threshold}$, is approximated as in \cite{hoangEffectDustRotational2021} as
\begin{equation} \label{eq:rsub}
    r_{\rm threshold} \simeq 155.3\left(\frac{L_{\rm bol}}{10^{6}L_{\odot}}\right)^{0.5}\left(\frac{T_{\rm threshold}}{1500\,{\rm K}}\right)^{-5.6/2} ~~~{\rm au},
\end{equation}
We observe that stellar radiation decreases as $r^{-2}$ for $r<r_{\rm threshold}$, but is more greatly attenuated due to the additional optical depth for $r\geq r_{\rm threshold}$.

When the temperature is below this threshold, there is a thin dust shell. This layer is irradiated by stellar radiation and plays an important role in heating the gas and dust in the outer zone. Therefore we also consider the contribution of the radiation from a thin dust shell as
\begin{equation} \label{eq:Ushell}
    \begin{split}
    U(T_{\rm shell}) &= \frac{\int^{20\,\rm \mu m}_{0.091\,\rm \mu m} L_{\lambda}(T_{\rm shell})e^{-\tau_{\lambda}}d\lambda}{4\pi r^{2}c u_{\rm ISRF}} \\
    &= \frac{\int^{20\,\rm \mu m}_{0.091\,\rm \mu m} 4\pi R^{2}_{\rm shell} \pi B_{\lambda}(T_{\rm shell})e^{-\tau_{\lambda}}d\lambda}{4\pi r^{2}c u_{\rm ISRF}} 
    \end{split},
\end{equation}
with $c$ the speed of light, 
where $T_{\rm shell}$ is calculated from $U(T_{\ast})(r=r_{\rm threshold})$ with an assumption that the hot shell is thin enough ($d_{\rm shell}\ll 1$) to prevent any radiative processes from occurring within it, resulting in the constant temperature of the shell, i.e.~$T_{\rm shell}(R_{\rm shell}) = T_{\rm shell}(R_{\rm shell}+d_{\rm shell})$.

At distances greater than $r_{\rm threshold}$, the radiation intensity is given by the combined contribution of these two radiation sources, and the corresponding dust temperature is calculated under the assumption of thermal equilibrium, with a dependence on the grain composition \citep[chapter 24]{drainePhysicsInterstellarIntergalactic2011},
\begin{equation} \label{eq:Tdust_r}
    \begin{split}
        T^{\rm sil}_{\rm dust}(r) &= 16.4\,{\rm K}\times \left[U(T_{\ast}) + U(T_{\rm shell})\right]^{1/6} ~~~~~ {\rm silicate}, \\
        T^{\rm car}_{\rm dust}(r) &= 19.5\,{\rm K}\times \left[U(T_{\ast}) + U(T_{\rm shell})\right]^{1/5.6} ~~~ {\rm carbonaceous},\\
        T^{\rm mix}_{\rm dust}(r) &= \left[0.625\left(T^{\rm sil}_{\rm dust}\right)^{4} + 0.375\left(T^{\rm car}_{\rm dust}\right)^{4}\right]^{1/4} ~~~{\rm mixture}.
    \end{split}
\end{equation}

The parameters related to the luminous source and the gas volume density are needed, namely the bolometric luminosity of the source ($L_{\ast}$) and its surface temperature ($T_{\ast}$), as well as $n_{0}$, $r_{0}$, and the slope $\alpha$. These temperatures represent the maximum values reached at a given distance, and the evolution from the initial temperature to this maximum is described by \Cref{eq:Td_hotcore} and the parameters are available
in \Cref{tab:protostellarab}.

For the one-dimensional treatment, we need to pay particular attention to the effectiveness and role of photoreactions, which are essentially determined by the visual extinction as
computed before.
In order to account for the self and mutual shielding of \ch{H2} and \ch{CO}, we use the treatment
from \citet{vandishoeckPhotodissociationChemistryInterstellar1988, federmanAtomicMolecularHydrogen1979} for both
the internal and external radiation field.

\begin{figure*}
    \centering
    \includegraphics[width=1.0\linewidth]{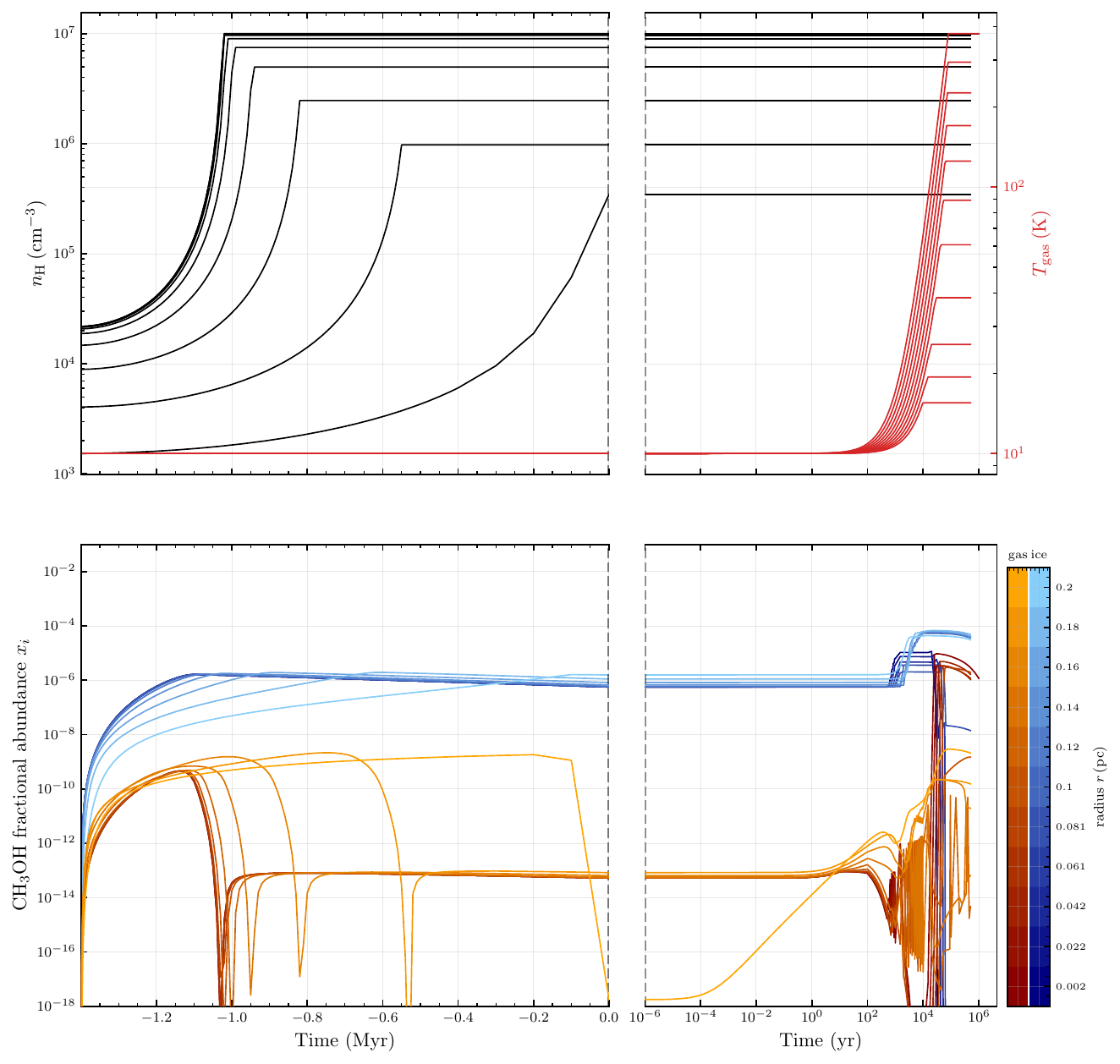}
    \caption{Density and temperature  profiles of a one-dimensional collapse
    and protostellar model, with the abundance of methanol in the gas phase and ice phase shown below.
    The model was run with $L_{\ast}=10^5\,L_{\odot}$, $r_{\rm 0}=5.0\times 10^{-2}\,\rm pc$, $n_0=10^7\,\rm cm^{-3}$.}
    \label{fig:onedimprofiles}
\end{figure*}

An example of a one-dimensional free-fall collapse and a subsequent
protostellar core model with an 
internal luminosity source of $L_*=10^5L_\odot$ is shown in \Cref{fig:onedimprofiles}.

\subsection{Chemical post-processing of hydrodynamical simulations}
\begin{figure}
    \centering
    \includegraphics[width=1.0\linewidth]{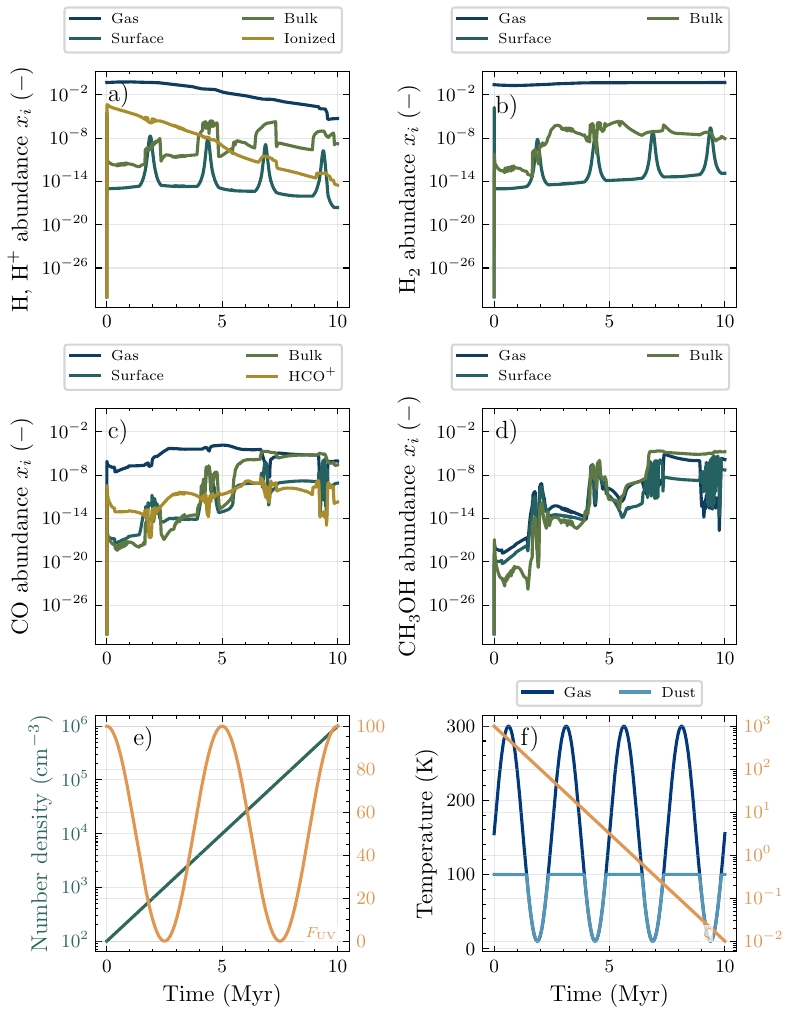}
    \caption{Example of input into the post-processing interface. The density is
    increasing from $10^2$ to $10^6$~cm$^{-3}$,
    the radiation field follows the function $F_\mathrm{UV}=50(1+\cos(2\pi t/5\mathrm{Myr}))$~Habing, the gas temperature fluctuates between $T_{\mathrm{gas}}=155 + 145 \sin(2\pi t/2.5\mathrm{Myr})$~K and the dust temperature is coupled with a maximum: $T_{\mathrm{dust}}=\min(T_{\mathrm{gas}},100)$~K. In $\text{a)-d)}$\,the evolution of
    hydrogen, molecular hydrogen, carbon monoxide, \ch{HCO+} and methanol are shown.}
    \label{fig:neath_example}
\end{figure}

\begin{figure}
    \centering
    \includegraphics[width=1.0\linewidth]{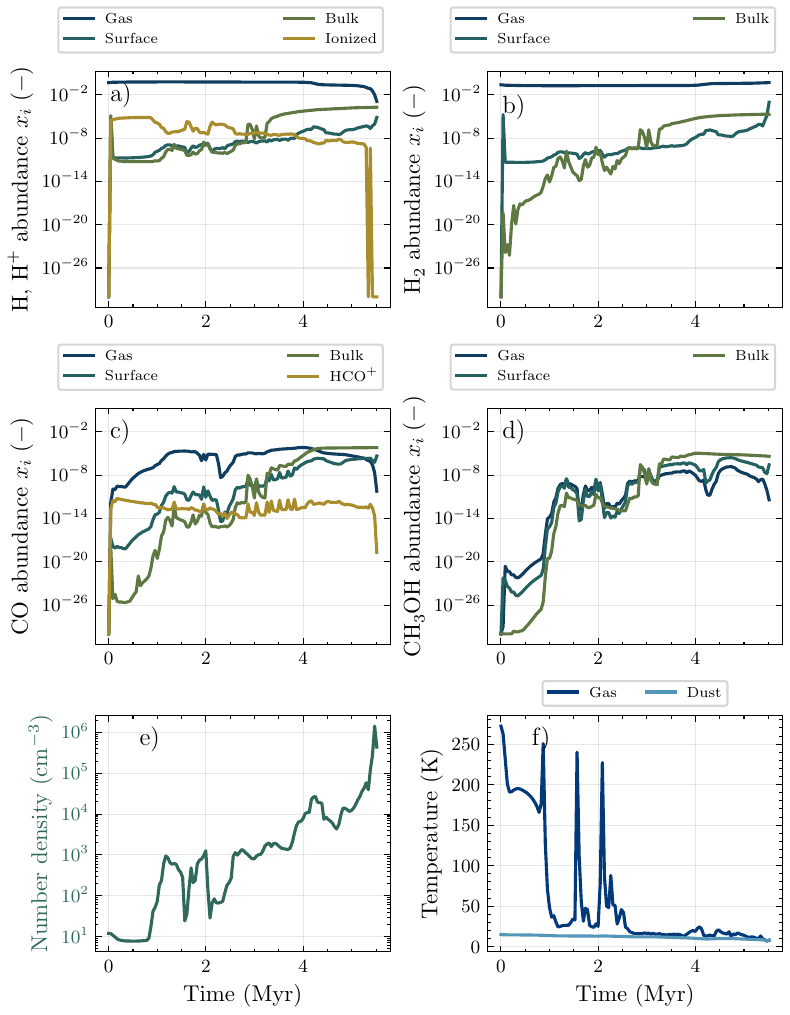}
    \caption{The chemical and physical evolution of a single particle from \texttt{NEATH} as computed directly with \texttt{UCLCHEM}. 
    The CRIR is set to the default value $\zeta=\zeta_0$ and
    the radiation field $F_\mathrm{UV}=1\;\mathrm{Habing}$.}
    \label{fig:neath_one_trajectory}
\end{figure}

\begin{figure}
    \centering
    \includegraphics[width=1.0\linewidth]{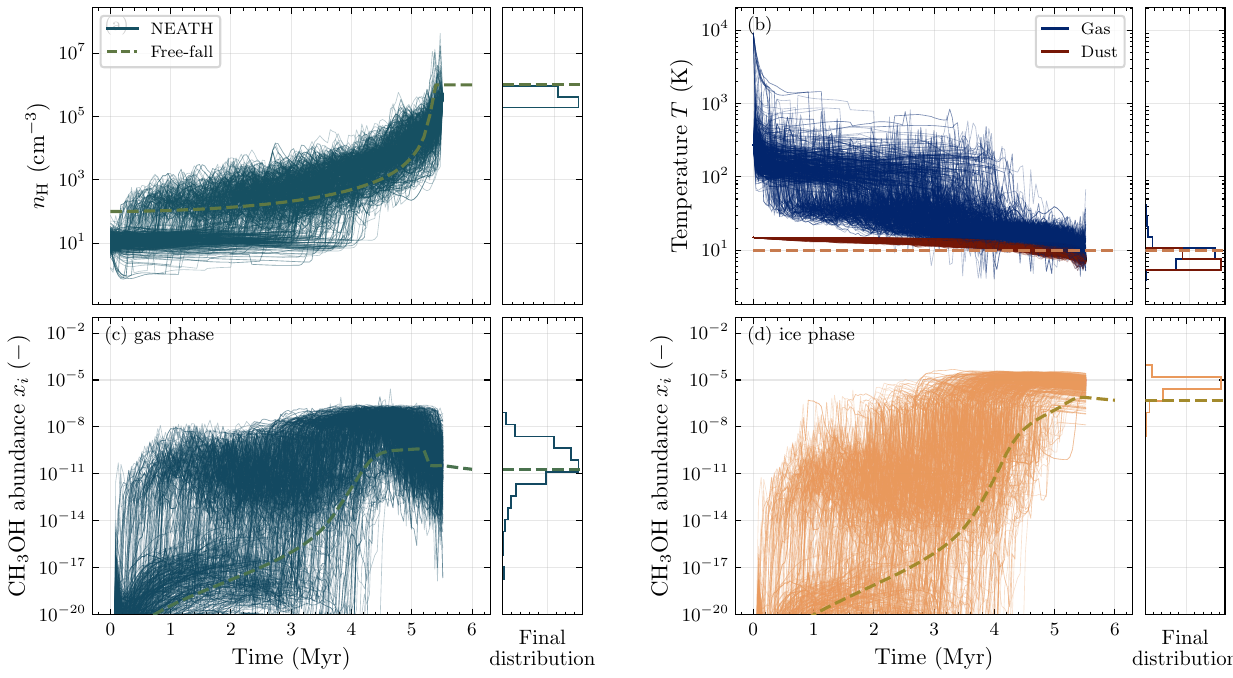}
    \caption{A comparison between more than 500 NEATH tracer particles and a typical \texttt{UCLCHEM} free-fall collapse model.}
    \label{fig:neath_methanol_comparison}
\end{figure}

Rather than assuming the parametrized models, the framework
also allows for using the physical history and even basic
chemical parameters directly from hydrodynamical simulations. 
This effectively  simulates the chemistry along the history,
treating each time step as a short model with constant parameters.
An example of arbitrary functions for gas temperature, dust temperature,
cosmic ray ionization rate and radiation field is shown in \Cref{fig:neath_example}, highlighting the flexibility of being able
to define custom profiles for effortless experimentation with new models or custom datasets.
To avoid the significant computational expense of implementing astrochemical reaction networks directly into MHD simulations \citep[e.g.][]{ferrada-chamorroChemicalPostprocessingMagnetohydrodynamical2021,bovinoChemicalAnalysisPrestellar2021,tritsisNonidealMagnetohydrodynamicSimulations2022,jensen3DPhysicochemicalModel2026}, it is common to perform the chemical modelling in a post-processing step. While in principle the dynamical evolution of the gas depends on its chemistry, via its impact on the thermal behaviour of the gas, for many relevant astrophysical scenarios this effect can be modest \citep{gloverApproximationsModellingCO2012}, and modern MHD codes often include a simplified internal chemical model to capture the most important effects self-consistently \citep[e.g.][]{gongSimpleAccurateNetwork2017,hunterImpactGMCCollisions2023}.

The post-processing module in \texttt{UCLCHEM} is based on the method described in the \texttt{NEATH} simulation \citep{priestleyNonEquilibriumAbundancesTreated2023}, which uses Monte Carlo tracer particles \citep{genelFollowingFlowTracer2013} to record the physical evolution of gas parcels in MHD simulations performed using the {\sc arepo} moving-mesh code \citep{springelPurSiMuove2010}. Specifically, these \texttt{NEATH} simulations follow the formation of a molecular cloud from diffuse atomic initial conditions (10 cm$^{-3}$), including turbulence and magnetic fields.
However, the nature of the underlying simulation is unimportant, as long as a suitable way of tracking the evolution of Lagrangian gas parcels is available \citep[e.g.][]{clementAstrochemicalModelsInterstellar2023,panessaEvolutionHCOMolecular2023,komichiChemicalEvolutionMolecular2024}. 

The post-processing module requires as input the physical evolutionary trajectory 
of each tracer particle over the course of the simulation. The model requires 
gas temperatures and densities. It allows for the dust temperature as an optional parameter, if not provided,
it will assume coupled temperatures, as well as either a constant or time dependent
radiation field and cosmic ray ionization rate. Additionally, if visual extinction $A_\mathrm{V}$ and (shielding) column
densities  $N_{\ch{H}}$, $N_{\ch{H2}}$, $N_{\ch{CO}}$, and $N_{\ch{C}}$ are provided, 
the framework will override the internal column density and visual extinction computation.

\texttt{UCLCHEM} treats each tracer particle as an independent point model for the purposes of its chemical evolution. For each timestep, the physical properties are updated from their associated arrays, the chemistry is evolved from the current to the subsequent timestep, and the physical properties are updated again in preparation for the next network integration. Note that the physical properties are assumed to remain constant between timesteps, so it is important that the interval between timesteps is short enough to capture any physical changes in the system. For most applications in star formation, this means resolving the local free-fall time with at least a few timesteps \citep{priestleyNonEquilibriumAbundancesTreated2023}; as the free-fall time decreases with density, a given timestep interval effectively corresponds to a maximum density at which the chemistry is fully converged.

The output of one tracer particle with timesteps of $4.3\times10^5$ yr is highlighted in \Cref{fig:neath_one_trajectory}:
its density is increasing whilst the particle is cooling, allowing for species to freeze out effectively, hydrogen to be converted
into its molecular form and methanol to be formed on the grains. 
Since the radiation field and cosmic ray ionization rate were not evolved
during the simulation of the hydrodynamics, they were assumed to be constant. 
If we then combine 500 particles from the \texttt{NEATH} simulation, as seen
in \Cref{fig:neath_methanol_comparison},
we can see that they all follow this behaviour and compare well to
a free-fall collapse model.

\subsection{Heating and cooling: Decoupling the gas and dust temperature}

\begin{figure}
    \centering
    \includegraphics[width=1.0\linewidth]{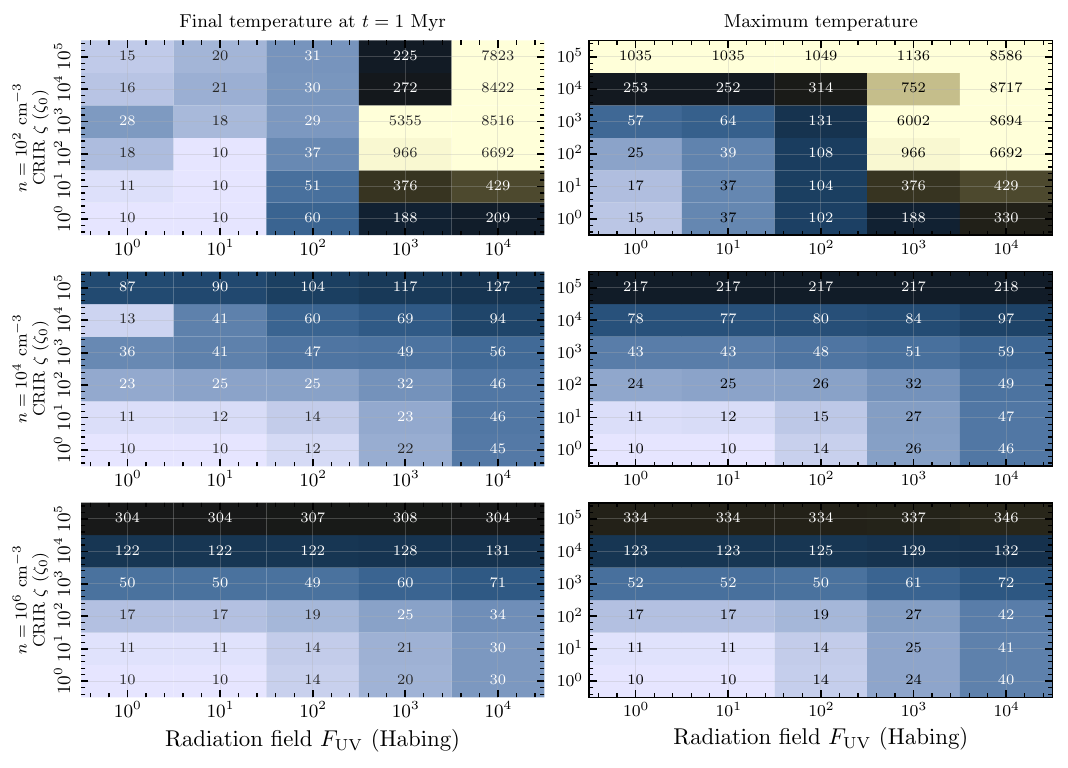}
    \caption{Cloud models with a constant number density, exposed to different Cosmic Ray Ionization Rates and Radiation fields, starting at an initial temperature of $T=10$K. The models evolve until one million years. The final temperature is shown on the left, the highest temperature is shown on the right. }
    \label{fig:temperature_comparison}
\end{figure}

\begin{figure*}
    \centering
    \includegraphics[width=1.0\linewidth]{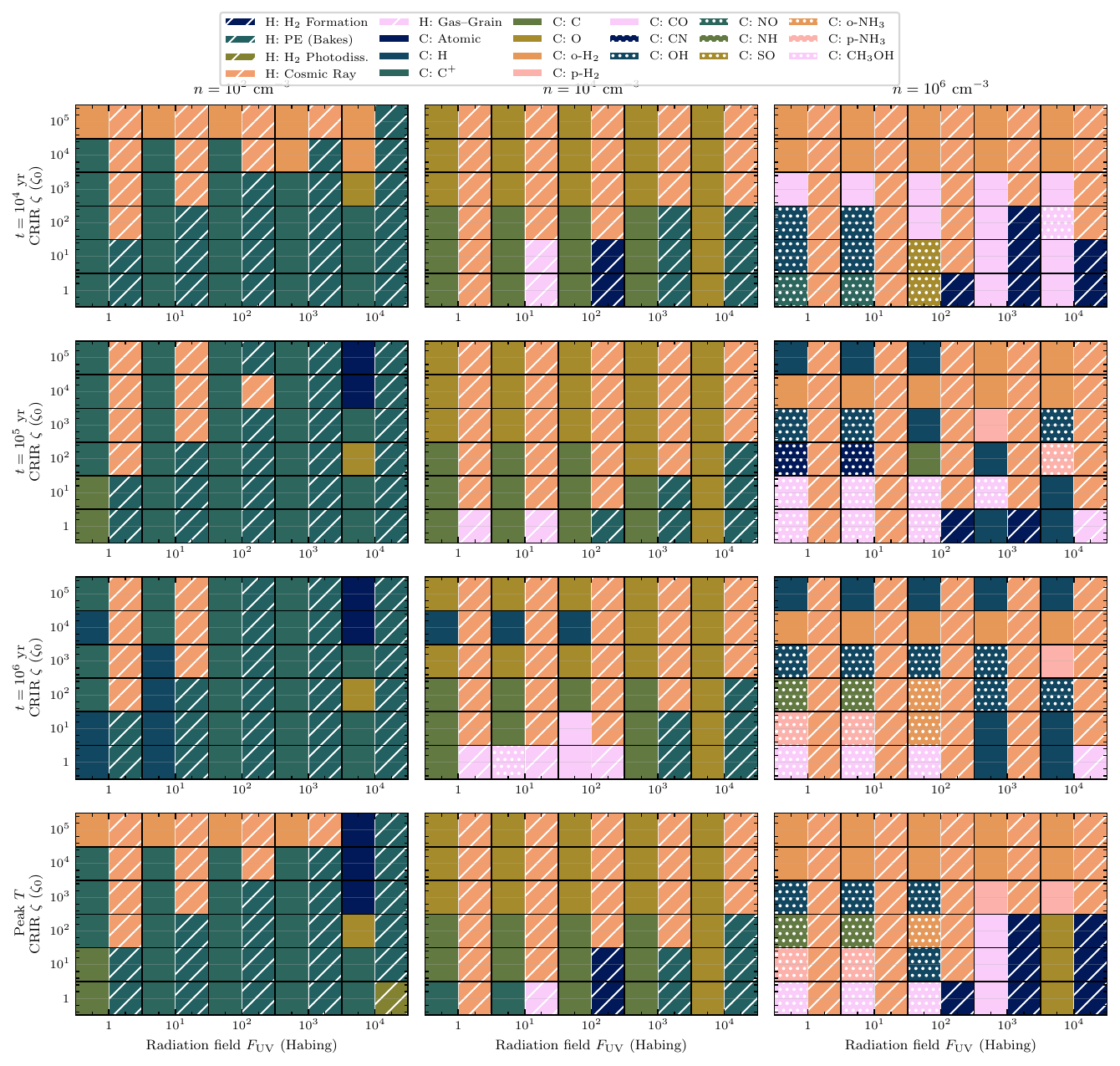}
    \caption{A comparison of the dominant heating and cooling mechanism for different times and at the time of the maximum temperature. Each model is started at $T=10\;\mathrm{K}$.}
    \label{fig:dominant_heating_mechanisms}
\end{figure*}

In order to account for feedback of different processes into the gas temperature, \texttt{UCLCHEM} follows its evolution taking into account the heating and cooling mechanisms which were first introduced into the framework by \cite{holdshipChemulatorFastAccurate2021}, following
\begin{equation}
\frac{\mathrm{d}T_{\rm gas}}{\mathrm{d}t} = \frac{\gamma_{\mathrm{ad}} -1}{k_\mathrm{B} n_{\rm H,nuclei}}\left(\sum_m \Gamma_m - \sum_m \Lambda_m \right),
\end{equation}
where $\Gamma$ and $\Lambda$ are the heating and cooling rate per unit volume, respectively (in erg~cm$^{-3}$~s$^{-1}$, as for each individual mechanism below) and $\gamma_{ad}$ is the adiabatic index, computed as 
\begin{equation}
    \gamma_{\mathrm{ad}} = \frac{5\left(n_{\mathrm{H}} + n_{\mathrm{He}} + n_{e^-} + n_{\mathrm{H_2}}\right) + 2n_{\mathrm{H_2}}}{3\left(n_{\mathrm{H}} + n_{\mathrm{He}} + n_{e^-} + n_{\mathrm{H_2}}\right) + 2n_{\mathrm{H_2}}}.
\end{equation} The processes adopted in \texttt{UCLCHEM} are similar to \texttt{UCLPDR}. The following subsections describe the analytical sources of heating and cooling. A comparison of
the final and maximum gas temperature for a grid of models with different densities, cosmic ray ionization rates and radiation fields is shown in \Cref{fig:temperature_comparison}. The dominant heating and cooling mechanism for the models are shown in \Cref{fig:dominant_heating_mechanisms}, highlighting the importance of many of the different heating and cooling mechanisms for different model conditions and the nonlinear effects the interplay between chemistry and heating can have. We now highlight the individual heating and cooling  mechanisms as present in the current version of the framework.

\begin{figure}
    \centering
    \includegraphics[width=1.0\linewidth]{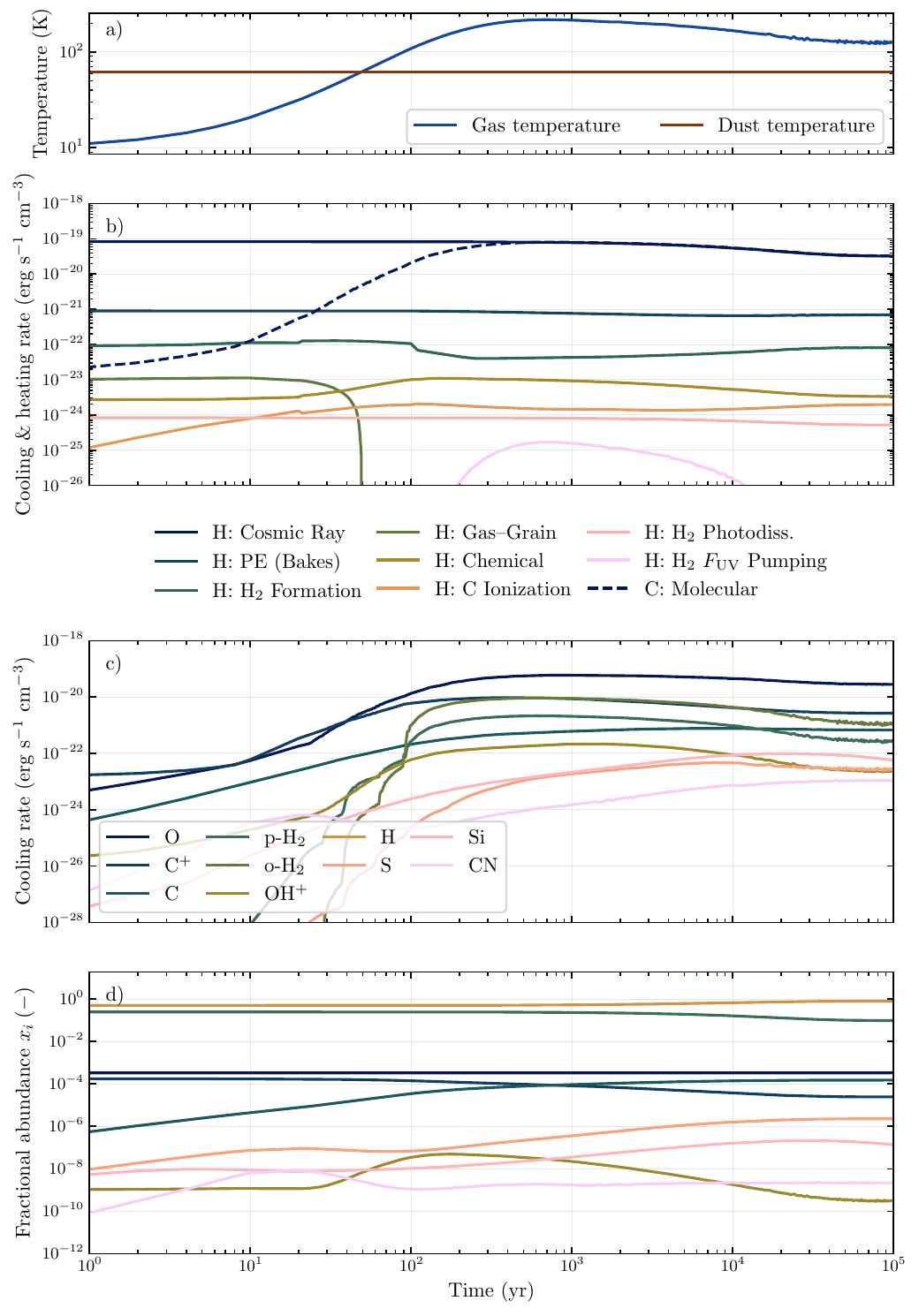}
    \caption{The evolution of a \texttt{UCLCHEM} model with $T_0=10\;\mathrm{K}$, $n_\mathrm{H,nuclei}=10^4\;\mathrm{cm}^{-3}$, $\zeta=10^5\zeta_0$, and $F_\mathrm{UV}=10^4\;\mathrm{Habing}$. Panel a) shows the temporal evolution
    of the gas and dust temperature. b) Shows the different heating and cooling mechanisms. 
    c) shows the individual contributions of the molecular cooling lines to the total
    molecular cooling and d) shows the coolant species fractional abundances.}
    \label{fig:heating_cooling_comparison}
\end{figure}

\subsubsection{Photoelectric heating and cooling of PAHs}
Photoelectric emission from dust grain surfaces serves as one of the primary heating mechanisms for the gas. We follow \cite{bakesPhotoelectricHeatingMechanism1994}, with an additional correction factor introduced in \cite{wolfireNeutralAtomicPhases2003} to calculate the heating rate for small grain as 
\begin{equation}
\Gamma_\mathrm{pe} = 1.30 \times 10^{-24}\,\epsilon\,F_\mathrm{UV}\,n_{\ch{H}} ~~~{\rm erg\,cm^{-3}\,s^{-1}},
\end{equation}
where the rate is determined by a parameter $\epsilon$ that scales both in low and high temperature regime
up to $10^4\;\mathrm{K}$,
\begin{equation}
\epsilon =
\frac{4.87 \times 10^{-2}}{1 + 4 \times 10^{-3}\,\left(\frac{F_\mathrm{UV} \sqrt{T_\mathrm{gas}}}{n_{\ch{e}} \Phi_\mathrm{PAH}}\right)^{0.73}} 
+ \frac{3.65 \times 10^{-2}\,(T_\mathrm{gas}/10^4\;\mathrm{K})^{0.7}}{1 + 2 \times 10^{-4}\,\frac{F_\mathrm{UV} \sqrt{T_\mathrm{gas}}}{n_{\ch{e}} \Phi_\mathrm{PAH}}},
\end{equation}
with $\phi_{\mathrm{PAH}}=0.4$.
At high temperatures, the photoelectric heating will compete with the recombination of said
electrons,
\begin{equation}
\Lambda_\mathrm{PAH} = 4.65 \times 10^{-30}\,
T_\mathrm{gas}^{\alpha}\,
\left(\frac{F_\mathrm{UV} \sqrt{T_\mathrm{gas}}}{n_{\ch{e}} \Phi_\mathrm{PAH}}\right)^{\beta}\,
n_{\ch{e}}\,
\Phi_\mathrm{PAH}\,
n_{\ch{H}}
\quad{\rm erg\,cm^{-3}\,s^{-1}}, \end{equation} where $\alpha=0.944$ and $ \beta = \frac{0.735}{T_\mathrm{gas}^{0.068}}$.

Alternatively, the mutually exclusive photoelectric treatment of \citet{weingartnerPhotoelectricEmissionInterstellar2001} is available:
\begin{equation}
\begin{split}
\Gamma_{\mathrm{pe}} &= 10^{-26} (F_\mathrm{UV} n_{\mathrm{H,nuclei}}) \times \\ &\frac{5.72 + 0.0345 T^{0.495}}{1 + 0.00708 \psi^{0.692} \left(1 + 0.0198 \psi ^{0.52}\right)}\quad{\rm erg\,cm^{-3}\,s^{-1}}.
\end{split}
\end{equation}
where $\psi=\frac{F_\mathrm{UV} \sqrt{T}}{n_{\ch{e}}}$.

\subsubsection{\ch{H2} formation and \ch{H2} dissociation treatments}
Since the formation of \ch{H2} is dependent on the temperature regime as
discussed in \Cref{sec:grainchem}, we introduce a temperature dependent piecewise treatment. Above the maximum grain temperature, we assume $1.5\,$eV of formation enthalpy will go to the kinetic energy, following \citet{hollenbachPhotodissociationRegionsInterstellar1999}. Below the maximum grain 
temperature, we instead use $0.1 + 4.2\,\eta$ eV per LH reaction \citep{hollenbachMoleculeFormationInfrared1979} and $0.6\,\eta$ eV per ER reaction \citep{lebourlotSurfaceChemistryInterstellar2012}:
\begin{equation}
\Gamma_{\ch{H2},\text{form}} = \begin{cases}
\begin{aligned}
&\left[0.1 + 4.2 \, \eta\right] R_{\ch{H2},\mathrm{LH}} \, n_{\ch{H},\text{nuclei}} \, x_{\#\ch{H}}^2 \\[0.5em]
&+ 0.6 \, \eta R_{\ch{H2},\mathrm{ER}} \, n_{\ch{H},\text{nuclei}}^2 \, \frac{x_{\ch{H}} \, x_{\#\ch{H}}} {x_{\mathrm{surface}}}
\end{aligned} & T < 150 \text{ K}, \\[1em]
1.5 \, R_{\ch{H2},\mathrm{CT}} \, n_{\ch{H},\text{nuclei}}^2 \, x_{\ch{H}} & T \geq 150 \text{ K},
\end{cases}
\end{equation}
where $x_\text{surface}$ is the total fractional surface density,
$R_{\ch{H2},\{LH,ER,CT\}}$ are the respective rates for the formation mechanisms and \begin{equation}
\eta(n, T) = \frac{n_{\ch{H},\text{nuclei}}}{n_{\ch{H},\text{nuclei}} + n_{\mathrm{cr}}(T)},
\end{equation}
defines a fraction of the excitation energy contributing to heating. The critical density
herein is defined as:
\begin{equation}
n_{\mathrm{cr}}(T_\mathrm{gas}) = \frac{10^6 \, T^{-1/2}}{1.6 \, x_{\ch{H}} \exp\left[-(400/T)^2\right] + 1.4 \, x_{\ch{H2}} \exp\left[-\frac{18100}{T+1200}\right]} \quad \mathrm{cm^{-3}}.
\end{equation}

For the photodissociation, $0.4\,$eV of kinetic energy per photodissociated molecule, the heating rate is:
\begin{equation}
    \Gamma_{\ch{H2},\rm photodis} = 0.4 \, {\rm eV} \, k_{\rm photodis} \, n_{\ch{H2}} ~~~{\rm erg\,cm^{-3}\,s^{-1}},
\end{equation}
where $k_{\rm photodis}$ is in s$^{-1}$
as treated in \citet{cazauxMolecularHydrogenFormation2002} and \citet{cazauxH2FormationGrain2004}.

\subsubsection{\ch{H2} Ultraviolet pumping}
The decay of \ch{H2} vibrational excited levels can heat the gas via collisional de-excitation. We follow \citet{hollenbachMoleculeFormationInfrared1979} and assume that the far-UV pumping rate is 9 times the photodissociation rate of \ch{H2} as 
\begin{equation}
\Gamma_{\rm FUV,pump} = \frac{2.2 \, {\rm eV} \times 9.0 \, k_{\rm photodis} \, n_{\ch{H2}}}{1 + n_{\rm cr}/n_{\ch{H},\rm nuclei}} ~~~{\rm erg\,cm^{-3}\,s^{-1}},
\end{equation}
where the critical density $n_{\rm cr}$ is calculated using the spontaneous emission Einstein coefficient with a rate of $A=10^{-6}\,\rm s^{-1}$, together with the collisional de-excitation coefficient for the $v=1\rightarrow 0$ transition with \ch{H2} as the collision partner and for the $v=2\rightarrow 0$ transition with \ch{H} as the collision partner.
Note that the the collisional de-excitation coefficient for $v=1\rightarrow 0$ with \ch{H2} is corrected in \citet{hollenbachMoleculeFormationInfrared1989}.

\subsubsection{Carbon photoionization heating}
Under intense UV radiation, neutral carbon may undergo photoionization, releasing photoelectrons that deposit on average about 1 eV of energy each, thereby heating the gas. The heating rate, accounting for the self-shielding of \textsc{C} and the mutual shielding of \ch{H}, is calculated as
\begin{equation}
    \Gamma_{\ch{C},\rm ion} = 1.0 \, {\rm eV} \times \alpha\, n_{\ch{C}}\, F_\mathrm{UV} f_{\rm shield}(\ch{C}) f_{\rm shield}(\ch{H2}) ~~~{\rm erg\,cm^{-3}\,s^{-1}},
\end{equation}
where $\alpha$ is the (s$^{-1}$) rate coefficient for the \ch{C} photoionization reaction, $f_{\rm shield}(\ch{CI})=\exp(-N(\ch{C})\alpha_{\ch{C}})$ with $\alpha_{\ch{C}}=1.1\times 10^{-17}\;\mathrm{cm}^2$ the ionization cross-section of \ch{C}, and $f_{\rm shield}(\ch{H2}) = \exp\{-\left[0.9 T^{0.27}_{\rm gas} (N_{\ch{H2}}/1.59\times 10^{21})^{0.45})\right]\}$.
\subsubsection{\ch{H2} Cosmic Ray ionization heating}
Following \citet{goldsmithMolecularDepletionThermal2001}, we calculate the volumetric heating rate due to \ch{H2} ionization, adopting a heating yield of 16$\,$eV per \ch{H2} ionization as
\begin{equation}
\Gamma_{\rm CR} = 16.0 \, {\rm eV} \times \zeta \times n_{\ch{H2}} ~~~ {\rm erg\,cm^{-3}\,s^{-1}},%
\end{equation}
\subsubsection{Turbulent heating}
Following \citet{blackHeatingCoolingInterstellar1987,rodriguez-fernandezWarmGalacticCenter2001,bisbas3DPDRNewThreedimensional2012}, the heating source due to the turbulent heating is
\begin{equation}
    \Gamma_{\rm turb} = 3.5\times 10^{-28}\left(\frac{v_{\rm turb}}{10^{5}\rm cm\,s^{-1}}\right)^{3}\left(\frac{1\,\rm pc}{L_{\rm turb}}\right)n_{\ch{H}} ~~~ {\rm erg\,cm^{-3}\,s^{-1}},
\end{equation}
where $v_{\rm turb}\,\rm (cm\,s^{-1})$ is the turbulent velocity and $L\,\rm (pc)$ is the turbulent lengthscale. By default, $v_{\rm turb}=1\,\rm km\,s^{-1}$ and $l_{\rm turb}=5\,\rm pc$ are set as in \citet{bisbas3DPDRNewThreedimensional2012}.
\subsubsection{Gas-grain collision heat exchange}
Collisions between gas particles and dust grains exchange thermal energy. This process heats the gas when $T_{\rm dust} > T_{\rm gas}$ and cools it otherwise. The efficiency depends on an accommodation coefficient. Following \citet{burkeGasgrainInteractionInterstellar1983}:
\begin{equation}
\Gamma_{\rm gg} = n_{\rm grain} \, \sigma_{\rm grain} \, n_{\ch{H}} \, \sqrt{\frac{8 k_{\rm B} T_{\rm gas}}{\pi m_{\ch{H}}}} \, \bar{\alpha}_{\rm acc} \, 2 k_{\rm B} (T_{\rm dust} - T_{\rm gas}) ~~~{\rm erg\,cm^{-3}\,s^{-1}},
\end{equation}
where $\bar{\alpha}_{\rm acc}$ is the ``average'' accommodation factor, and $\sigma_{\rm grain}$ is the grain cross-section.

\subsubsection{Atomic line cooling}
We utilize the cooling rates for the collisional excitation cooling of \ch{H} and \ch{He+}, collisional ionization of \ch{H}, \ch{He}, \ch{He+}, and dielectric recombination cooling of \ch{He+} \citep{cenHYDRODYNAMICAPPROACHCOSMOLOGY}:
\begin{equation}
\begin{split}
\Lambda_{\rm atomic} = & \, 7.5 \times 10^{-19} \, \left(1+\sqrt{T_{5}}\right)^{-1} \, e^{-118348/T} \, n_{\ch{e-}} \, n_{\ch{H}} \\
& + 5.54 \times 10^{-17} \, T^{-0.397} \, \left(1+\sqrt{T_{5}}\right)^{-1} \, e^{-473638/T} \, n_{\ch{e-}} \, n_{\ch{He+}} \\
& + 1.27 \times 10^{-21} \, \sqrt{T} \, e^{-157809.1/T} \, \left(1+\sqrt{T_{5}}\right)^{-1} \, n_{\ch{e-}} \, n_{\ch{H}} \\
& + 9.38 \times 10^{-22} \, \sqrt{T} \, e^{-285335.4/T} \, \left(1+\sqrt{T_{5}}\right)^{-1} \, n_{\ch{e-}} \, n_{\ch{He}} \\
& + 4.95 \times 10^{-22} \, \sqrt{T} \, e^{-631515/T} \, \left(1+\sqrt{T_{5}}\right)^{-1} \, n_{\ch{e-}} \, n_{\ch{He+}} \\
& + 1.24 \times 10^{-13} \, T^{-1.5} \, e^{-470000/T} \, (1 + 0.3 \, e^{-94000/T}) \, n_{\ch{e-}} \, n_{\ch{He+}},
\end{split}
\end{equation}
where $T_5 = T/(10^5\;\mathrm{K})$, $T_3 = T/(10^3\;\mathrm{K})$ and $T_6 = T/(10^6\;\mathrm{K})$ (with $T$ in K, so $T_5$, $T_3$, $T_6$ are dimensionless).
At high temperatures,
above $T=10^5\;\mathrm{K}$, we additionally include the recombination cooling of \ch{H+} and \ch{He+} as
well as the Bremsstrahlung cooling of the ions: 
\begin{equation}
\begin{split}
\Lambda_{\rm atomic,T>10^5\;\mathrm{K}}=&\,\Lambda_{\rm atomic}+ \, 8.7 \times 10^{-27} \, \sqrt{T} \, (T_3)^{-0.2} \, \frac{n_{\ch{e-}} \, n_{\ch{H+}}}{1 + T_6^{0.7}} \\
& + 1.55 \times 10^{-26} \, T^{0.3647} \, n_{\ch{e-}} \, n_{\ch{He+}} \\
& + 1.42 \times 10^{-27} \, \sqrt{T} \, n_{\ch{e-}} \, (n_{\ch{He+}} + n_{\ch{H+}}) \, g_{\rm ff}.
\end{split}
\end{equation}
where the gaunt factor is computed using \citet{katzCosmologicalSimulationsTreeSPH1996}: $g_{\rm ff} = 1.1 + 0.34 \, \exp\left(-\frac{(5.5 - \log_{10} T)^2}{3}\right)$.
Effectively, these cooling rates only exceed nonzero for double floats above
$T=300$K.

\subsubsection{Collisional cooling}
Following \citet{hiranoRADIATIVECOOLINGIMPLEMENTATIONS2013} and \citet{ripamontiFragmentationFormationPrimordial2004} via \citet{grassiKROMEPackageEmbed2014}:
\begin{equation}
\Lambda_{\rm CIE} = 10^{f(T)} \times \frac{1 - e^{-\tau}}{\tau},
\end{equation}
where the optical depth parameter is:
\begin{equation}
\tau = \left(\frac{n_{\rm H_2}}{7 \times 10^{15}\;\mathrm{cm^{-3}}}\right)^{2.8},
\end{equation}

and the function is defined as:
\begin{equation}
f(T) = 
    \begin{cases}
3.0 \, \log_{10} T - 21.297 & T \geq 10^5 \;\mathrm{K}\\
    \sum_{i=0}^{5} a_i \, (\log_{10} T)^i & 891\;\mathrm{K}\leq T < 10^5\;\mathrm{K}\\
    \sum_{i=0}^{5} b_i \, (\log_{10} T)^i &\;100\;\mathrm{K}\leq T <891\;\mathrm{K}\\
\end{cases},
\end{equation}
where $a = [-180.99, 168.47, -67.50, 13.51, -1.320, 0.0500]$ and
$b = [-30.33, 19.00, -17.15, 9.495, -2.548, 0.2654]$.

\subsubsection{Continuum emission cooling}
The continuum emission from dust in the Rayleigh-Jeans regime can cool the dust (\citealt{hiranoRADIATIVECOOLINGIMPLEMENTATIONS2013} and \citealt{ripamontiFragmentationFormationPrimordial2004} via \citealt{grassiKROMEPackageEmbed2014}):
\begin{equation}
\Lambda_{\rm cont} = 4 \sigma_{\rm SB} \, T^4 \, \kappa \, \rho_{\rm gas} \, \min\left[\tau^{-2}, 1\right],
\end{equation}
$\sigma_{\rm SB}$ is the Stefan-Boltzmann constant,
$\rho_{\rm gas} = \min(0.5, n_{\ch{H}} \, m_{\ch{H}} \times 1.22)\;\mathrm{g\,cm^{-3}}$ is the mass density (assuming mean molecular weight 1.22),
$\kappa = 10^{1.000042 \, \log_{10} \rho_{\rm gas} + 2.14989}\;\mathrm{cm^2\,g^{-1}}$ is the Lenzuni opacity fit and
 $\tau = \sqrt{\frac{\pi k_{\rm B} T}{\rho_{\rm gas} \, m_{\ch{H}} \times 1.22 \times G}} \, \kappa \, \rho_{\rm gas}$ is the optical depth, with $G=6.674\times10^{-8}\;\mathrm{cm^3\,g^{-1}\,s^{-2}}$ the gravitational constant (cgs units).

\subsubsection{\ch{H2} vibrational cooling}
We treat the vibrationally excited levels of \ch{H2} as a single pseudo level
with effective rates of spontaneous emission, collisional excitation,
$F_{\rm UV}$ pumping and photodissociation that describe the behaviour of all the vibrational levels combined:
\begin{equation}
    \Lambda_{\ch{H2},\rm vib} = k_{\rm B} \, \Delta E_{10} \, C_{10} \, n_{\ch{H}} \, e^{-\Delta E_{10}/T} \, n_{\ch{H2}} \, \frac{A_{10} + k_{\rm photodis}}{C_{10} \, n_{\ch{H}} + A_{10} + k_{\rm photodis}},
\end{equation}
With the effective heating from $F_{\rm UV}$ pumping: 
\begin{equation}
\Gamma_{\ch{H2},\rm vib} = n_{\ch{H2}} \, n_{\ch{H}} \, \frac{R_{\rm pump,eff} \, k_{\rm B} \, \Delta E_{\rm eff}}{1 + \frac{A_{10} + R_{\rm photo,eff}}{C_{10} \, n_{\ch{H}}}},
\end{equation}
where $\Delta E_{10} = 6587$ K is the energy gap between $v=1$ and $v=0$, $A_{10} = 8.6 \times 10^{-7}$ s$^{-1}$ the Einstein A-coefficient, $C_{10} = 5.4 \times 10^{-13} \sqrt{T}$ cm$^3$ s$^{-1}$ is the collisional rate coefficient, $\Delta E_{\rm eff} = 23500$ K the characteristic vibrational energy, $R_{\rm pump,eff} = 11.2 \times k_{\rm photodis}$ the effective vibrational pumping and  $R_{\rm photo,eff} = 18.0 \times k_{\rm photodis}$ the effective photodissociation rate respectively.

\subsubsection{Molecular line emission cooling}
In order to account for the cooling due to line emission of molecules, 
the level population of the molecules must be approximated, providing us with
the cooling rates. \texttt{UCLCHEM} adopts the radiative transfer computation from 
UCLPDR, which was iteratively improved to include more molecules over time. It follows closely the treatment
of \cite{dejongHydrostaticModelsMolecular1980} and the process of computing the line populations and cooling is described in detail in \cite{vandertakComputerProgramFast2007}. 
We include a range of molecules up to \ch{CH3OH}, listed in
\Cref{tab:coolants}. By default, only the default species from \texttt{UCLPDR} are
used as a coolant: $\ch{H, C^+, C, O, o-H2, p-H2, CO}$. However,
as can be seen in \Cref{fig:dominant_heating_mechanisms}, the line cooling contributions
of non default species can often also contribute meaningfully. \Cref{fig:heating_cooling_comparison} shows the
contributions of each of the heating and cooling mechanisms, 
as well as the individual line cooling contributions and abundances.
For the sake of computational speed and memory efficiency, the level population matrices are implemented using custom sparse arrays and retained for each parcel, allowing the solver to more easily solve the next timestep, 
given a small change in physical conditions and temperatures. If the
density changes, this cached level population state is scaled 
proportionally to the change in density.
Alternatively, the user can choose to reset the level populations to the ground
level or a local thermal equilibrium approximation at each solver timestep. 

\begin{figure*}
    \centering
    \includegraphics[width=1.0\linewidth]{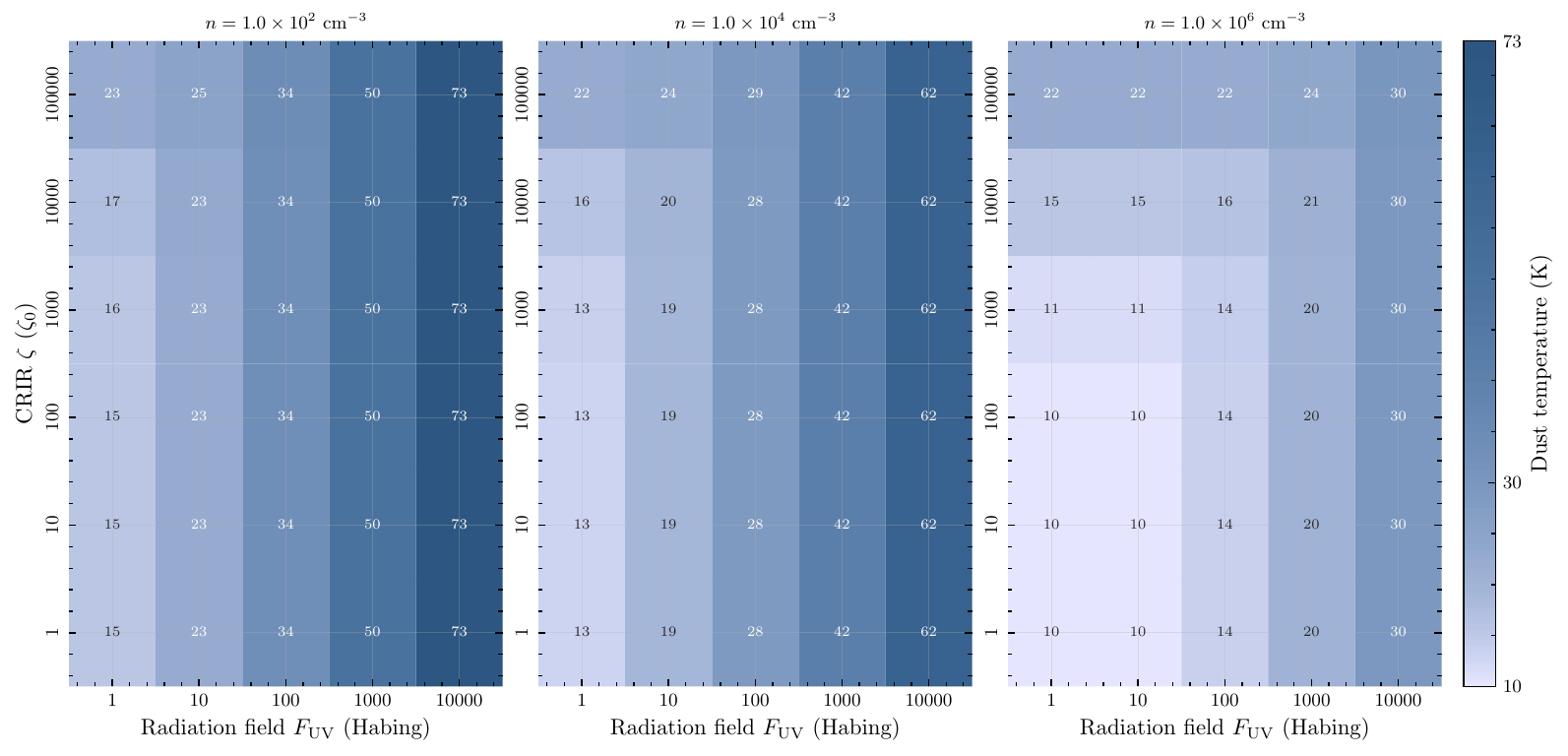}
    \caption{A comparison of the dust temperature with heating induced by different Cosmic Ray Ionization Rates and Radiation fields. The dust temperature is assumed to be constant and independent of the gas temperature.}
    \label{fig:dusttemp_comparison}
\end{figure*}

\begin{table}
\centering
\caption{Chemical species included in the atomic line cooling mechanism, with number of energy levels and energy range (in cm$^{-1}$) from the LAMDA data files. The default coolants are indicated with a $\dagger$.}
\begin{tabular}{lrrr}
\hline\hline
Species & $N_\mathrm{levels}$ & $E_\mathrm{min}$ (cm$^{-1}$) & $E_\mathrm{max}$ (cm$^{-1}$) \\
\hline
\ch{H}$^\dagger$                  &   2 &     0.00 & 82258.92 \\
\ch{C+}$^\dagger$                 &   2 &     0.00 &    63.40 \\
\ch{C}$^\dagger$                  &   3 &     0.00 &    43.41 \\
\ch{O}$^\dagger$                  &   3 &     0.00 &   226.99 \\
\ch{N+}                         &   3 &     0.00 &   130.80 \\
\ch{S}                          &   3 &     0.00 &   573.64 \\
\ch{Si}                         &   3 &     0.00 &   223.16 \\
$o$-\ch{H2}$^\dagger$             &  25 &   118.50 & 14495.46 \\
$p$-\ch{H2}$^\dagger$             &  28 &     0.00 & 15228.88 \\
\ch{CO}$^\dagger$                 &  41 &     0.00 &  3136.51 \\
\ch{CN}                         &  41 &     0.00 &   793.20 \\
\ch{CS}                         &  41 &     0.00 &  1336.42 \\
\ch{OH}                         &  20 &     0.00 &   608.20 \\
\ch{NO}                         & 100 &     0.00 &   202.30 \\
\ch{O2}                         &  48 &     0.00 &  1422.50 \\
\ch{CH+}                        &  14 &     0.00 &  2490.49 \\
\ch{HCO+}                       &  22 &     0.00 &   686.64 \\
\ch{HCS+}                       &  31 &     0.00 &   661.29 \\
\ch{N2H+}                       &  31 &     0.00 &  1442.66 \\
\ch{NO+}                        &  19 &     0.00 &   679.22 \\
\ch{NS+}                        &  28 &     0.00 &   631.03 \\
\ch{OH+}                        &  49 &     0.00 &  1174.16 \\
\ch{NH}                         &  25 &     0.00 &  1168.50 \\
\ch{HCl}                        &  40 &     0.00 &  1142.05 \\
\ch{HCN}                        &  26 &     0.00 &   959.62 \\
\ch{HNC}                        &  26 &     0.00 &   981.47 \\
\ch{HNCO}                       &  68 &     0.00 &   144.74 \\
\ch{HC3N}                       &  21 &     0.00 &    63.73 \\
\ch{SiO}                        &  41 &     0.00 &  1185.06 \\
\ch{SiS}                        &  41 &     0.00 &   496.04 \\
\ch{SO}                         &  91 &     0.00 &   675.84 \\
\ch{SO2}                        & 198 &     0.00 &   247.63 \\
\ch{OCS}                        &  99 &     0.00 &  1964.03 \\
$o$-\ch{H3O+}                   &   9 &     5.20 &   258.75 \\
$p$-\ch{H3O+}                   &  14 &     0.00 &   253.86 \\
$o$-\ch{NH3}                    &  22 &     0.00 &   553.60 \\
$p$-\ch{NH3}                    &  24 &    15.38 &   293.84 \\
$o$-\ch{CH3CN}                  &  52 &     0.00 &   214.00 \\
$p$-\ch{CH3CN}                  &  75 &     5.58 &   204.36 \\
$o$-\ch{C3H2}                   &  47 &     1.63 &    82.37 \\
$p$-\ch{C3H2}                   &  48 &     0.00 &    82.40 \\
\ch{SiC2}                       &  40 &     0.00 &    50.13 \\
\ch{CH3CN} (\ch{H2} collisions)       & 251 &     0.00 &   798.53 \\
\ch{CH3OH}                      & 256 &     0.00 &   966.96 \\
\hline
Total                           & 2071 & & \\
\hline
\end{tabular}
\label{tab:coolants}
\end{table}

\subsubsection{Chemical heating and cooling of gas-phase reactions}
In order to support the effect of reactions heating or cooling the system,
we optionally include this term in the framework, enabling the user to
custom-define both reaction-dependent changes in enthalpies as well as 
enthalpies of species to auto-compute each reaction.
It is important to note that this treatment is separate from the 
desorption probability on the ices, which also depends on the enthalpies. These enthalpies could be different, because in some gas-phase reactions absorption of a photon (or another external energy source) is involved.
The chemical heating and cooling contributions for each reaction
are obtained by multiplying the energy released by the reaction by its rate:
\begin{equation}
\Gamma_\text{chem}=\sum_j R_j \Delta H_j\quad{\rm erg\,cm^{-3}\,s^{-1}},
\end{equation}
where $R_j$ is the rate of reaction $j$ (in cm$^{-3}$~s$^{-1}$) and the energy released is the negative of the enthalpy of each reaction $j$, which can be expressed as
\begin{equation}
\Delta H_j = -\sum_{p} \Delta H_f^0 + \sum_{r} \Delta H_f^0\quad({\rm erg}).
\end{equation}
where $H_f^0$ are the enthalpies of formation at $T=0\;\rm K$, converted to erg at network-build time from whichever units the source data are given in (eV for the \citet{clavelChemicalThermalEquilibrium1978} values in \Cref{app:clavel}, kcal~mol$^{-1}$ for the directly-derived species enthalpies). This term is disabled by default. \texttt{UCLCHEM}
chooses to combine two sources for these enthalpies, recombination and ion-molecule reactions from \citet{clavelChemicalThermalEquilibrium1978}, as well
as reaction enthalpies derived directly from the
species. A list of the reaction enthalpies for 
the first source can be found in \Cref{app:clavel}. The second uses the enthalpies
that can be retrieved from several databases as reviewed in \citet{bovinoASTROCHEMICALMODELLINGPractical2023}, we include the enthalpies for
species from the default network. Another tool for computing the enthalpies is provided in \citet[Appendix B]{dijkhuisSensitivityAnalysisInterstellar}.

\subsubsection{Heating and cooling of the dust grains}
The temperature of the dust grain is parametrized and assumed to be at equilibrium. Three treatments are available: a UV-photon-based method from \citet{hollenbachLowDensityPhotodissociationRegions1991}, a visual-extinction-based method \citep{hocukParameterizingInterstellarDust2017} and a
version that additionally accounts for cosmic ray heating \citep{ivlevGasDustTemperature2019}.
The dust temperature is limited to a range, which is set to $T=[10,10^3]\rm K$ by default.

The first method follows \citet[equation 5]{hollenbachLowDensityPhotodissociationRegions1991}:
\begin{align}
T_{\rm dust,HHT} &= \Big(8.9 \times 10^{-11}\nu_0F_\mathrm{UV} + T_{\mathrm{CMB}}^{5}+\\
&{0.0345}\tau_{100}\big(0.42 - \ln\left(0.0345\tau_{100}T_0\right)\big)T_0^5\Big)^{0.2},
\end{align}
with $\tau_{100} = 10^{-3}$ (fixed), $\nu_0=2.65\times10^{15}\mathrm{s}^{-1}$ and $T_0 = 12.2F_\mathrm{UV}^{0.2}$.

The second method instead uses a parametrization based on semi-analytical solutions of the dust thermal balance and a collection of observational measurements \citep{hocukParameterizingInterstellarDust2017}:
\begin{equation}
    T_{\rm dust,Hocuk} = \left[11.0 + 5.7 \, \tanh(0.61 - \log_{10} A_\mathrm V)\right] \times (1.7 \, F_\mathrm{UV})^{1/5.9}.
\end{equation}
The last method extends the second by including a correction due to cosmic rays \citep{ivlevGasDustTemperature2019}:
\begin{equation}
    T_{\rm dust, Ivlev} = T_{d,0}\left[1+0.202\left(\frac{\zeta}{10^{-16}\,\rm s^{-1}}\right)\left(\frac{T_{\rm dust,Hocuk}}{6}\right)^{-6}\right]^{1/6},
\end{equation}
where $T_{d,0}=T_{\rm dust,Hocuk}$ is the base dust temperature computed above.

Of these three, we adopt the Ivlev method by default, as it accounts for both $\zeta$ and $A_\mathrm{V}$. 
This is especially relevant for models with a high $\zeta$, where the dust temperature
at lower radiation fields can be increased from 10 to 22$\,$K at higher densities. A comparison across a range of densities, $\zeta$ and $F_\mathrm{UV}$ is shown in \Cref{fig:dusttemp_comparison}.

\section{The \texttt{UCLCHEM} Framework}
\label{sec:uclchem}
The \texttt{UCLCHEM} framework consists of different modules, each focusing on a specific aspect of the astrochemical modelling.
So far, we have divided the components of \texttt{UCLCHEM} into two broad categories, physics and chemistry. These two categories are then implemented via various modules and subroutines in \texttt{UCLCHEM}. The core functionality in \texttt{UCLCHEM} is solving the differential equations of the chemistry; providing the user with the abundances of each species as a function over time and physical conditions. The chemical state can then be carried
over from physical model to physical model, allowing the user to describe increasingly complex astrophysical histories. 

The aforementioned modules are all implemented in 
Fortran and used to be exposed via a \texttt{main.f90} interface, relying on configuration files and tabular
storage for everything. With the advent of modern
\texttt{UCLCHEM} as presented in \citet{holdshipUCLCHEMGasgrainChemical2017},
a Python interface was introduced, using the \texttt{f2py} module from the \texttt{numpy} ecosystem \citep{petersonF2PYToolConnecting2009,harris2020array}. 
This first \texttt{UCLCHEM} wrapper
implemented a functional programming interface 
with many Fortran routines simply exposed and little interactivity. 
With this release of \texttt{UCLCHEM}, we improve this by introducing a 
new Object Oriented Programming (OOP) interface, keeping the old
functional interface for legacy purposes. With the new and improved OOP 
interface, the user can more easily run \texttt{UCLCHEM} in interactive
programming sessions, edit the Fortran parameters at runtime and integrate
\texttt{UCLCHEM} into statistical inference and sensitivity study pipelines such as \citet{keilUCLCHEMCMCMCMCInference2022,heylStatisticalMachineLearning,dijkhuisSensitivityAnalysisInterstellar}.
This section presents some of the core programmatic modules of \texttt{UCLCHEM},
some numerical considerations of solving the underlying numerical equations and
a more extensive explanation of the OOP interface.

\subsection{\texttt{Makerates}: building astrochemical reaction networks}
Chemical networks need to be composed before running \texttt{UCLCHEM}. For this purpose the \texttt{Makerates} tool was introduced and allows the user not only to easily create chemical reactions networks, but also to
check the consistency of such reaction networks, and finally write down the terms of the differential equations
that describe them. 

The input of \texttt{Makerates} consists of one file listing the species and
an arbitrary amount of files with reactions. The species input file uniquely describes each species by its mass, binding energy, enthalpy of formation, and some optional others, such as its moments of inertia or diffusion barrier. The reaction files list the reactions, each with its reactants and products, a string denoting its reaction type for any reaction that is not a standard two-body Arrhenius-Kooij reaction, parameters for the reaction rate constants $\alpha$, $\beta$ and $\gamma$ and finally the valid temperature range of the rates for each individual reaction: $[T_{\min}, T_{\max}]$. A brief summary of each individual reaction mechanism, and reaction rate constants can be found in \Cref{tab:ratestable}. Typically, we choose one chemical gas phase reaction
database and one custom reactions file, defining reactions of specific
interest as well as reactions in the ice phase.

The user does not need to pre-define the reaction that leads each gas phase species to freeze to the surface and bulk of the grains manually, 
as \texttt{Makerates} will automatically create a freeze-out and desorption pathway. However, the user is free to define custom freeze-out or desorption
pathways, such as desorbing species to their ionized variant or 
desorbing into two dissociated products rather than the species itself.

For the reactions on the ices, the user has to define the reactions that
occur on the grain surfaces. If the user provides the LH and ER reactions on the grain, but they do not
provide the equivalent reaction leading to desorption into the gas phase,
\texttt{Makerates} will then automatically add these \texttt{LH} and \texttt{ER} desorption pathways as \texttt{LHDES} and \texttt{ERDES} respectively. 

A step by step description of the \texttt{Makerates} routine can then be
summarized as follows:
\begin{enumerate}
\item Inputs a list with all species $i$.
\item Inputs a list of reactions $j$ from the custom reactions file, as well as  UMIST or KIDA reactions based on what which
species are defined in the above step.
\item Checks that each of the species can freeze out onto the grains, making sure
the frozen out species exists.
\item Creates the bulk species and reactions, based on the surface grain reactions.
\item Add the desorption reactions, both thermal and non-thermal,
checking that the gas-phase desorption products exist in the network.
\item Optional: If there are excited species, add excited state surface reactions.
\item Optional: Derive the enthalpy change for each reaction for heating and cooling.
\item Checks the branching ratios for reactions on the grains, warning the user for any incorrect branching ratio.
\item Sort the species based on phase first, and then by increasing mass.
\item Perform a  final check that no species are present that have no reactions.
\end{enumerate}
This entire routine results in a \texttt{Network} Python object, with all species and
reactions stored as their own objects, `Species` and `Reaction` respectively.
If the user runs the \texttt{Makerates} routine from the command line interface, with
a configuration file, it will then by default write key files to the \texttt{UCLCHEM} Fortran directory: \texttt{f2py-constants.f90, network.f90} and \texttt{odes.f90}. The first file contains some essential parameters
such as the number of reactions and species, the second contains hard-coded
arrays that define the chemical network, and the third defines the right
hand side of the differential equation that we solve. 
Since these files are hard-written to the Fortran source code, the
\texttt{UCLCHEM} package must be reinstalled before the new network can be used.
The entire routine relies on the \texttt{Network} submodule in Python, defining
\texttt{Species} and \texttt{Reaction} as objects and allowing the user
to interact with them. This also allows the user to load the chemical
network at runtime, either to plot and visualize, or to build advanced chemical 
networks not easily obtained via the command line interface.

Alternatively, at runtime, the user can interact with the compiled
chemical network. Many parameters, such as $\alpha$, $\beta$ and $\gamma$ of each reaction, can be edited live before running a model. This allows the user
to perform parameter sensitivity studies \citep{grassiComplexityReductionAstrochemical2012,dijkhuisSensitivityAnalysisInterstellar, vandesandeSenseSensitivityUncertainty2026} of core network parameters without
having to reinstall the package. An important note here is that removing reactions by setting their rates to zero is trivial, but adding reactions at runtime is not possible; so for sensitivity and network studies, the user must
define their largest possible superset of chemical reactions and species.

\subsection{Solving astrochemical differential equations}

\begin{figure}
    \centering
    \includegraphics[width=1.0\linewidth]{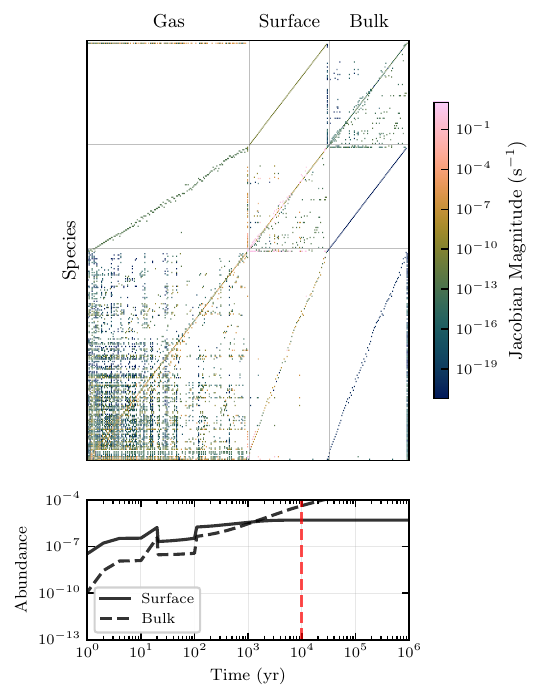}
    \caption{The Jacobian of a model at an isothermal constant density $n_\mathrm{H}=10^4$~cm$^{-3}$ and $T=75$~K at $10^4$~years.
    The figure clearly highlights the gas, surface and bulk species with the triple diagonal band structure. The
    offset-diagonal below the main diagonal shows important mechanisms such as freeze-out, thermal desorption and the ice
    swapping mechanisms. The bottom plot shows the evolution of the surface and bulk abundances, with the bulk becoming the reservoir of the ice species after $t=10^3\;\mathrm{yr}$.}
    \label{fig:jacobian}
\end{figure}

\begin{figure}
    \centering
    \includegraphics[width=1.0\linewidth]{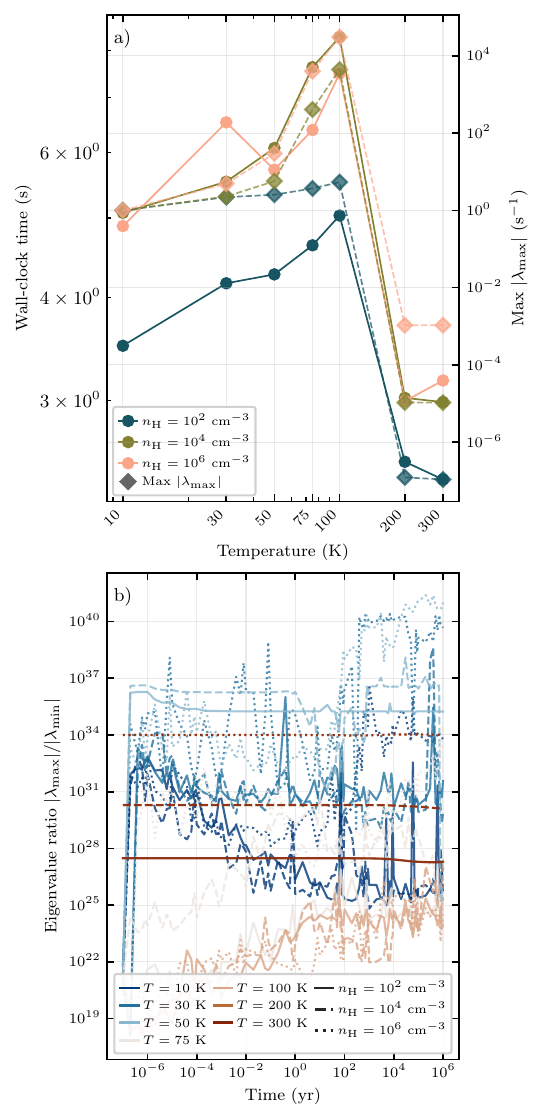}
    \caption{a) Shows the average wall clock time of 10 constant density and temperature models evaluated on a AMD EPYC\textsuperscript{TM} 7702P. b) Shows the
    eigenvalue ratio as a proxy for the stiffness of the individual models over time. }
    \label{fig:stiffness}
\end{figure}

The core solver routine at the heart of \texttt{UCLCHEM} solves  the differential equation for each chemical species' number density
as a function of time. Since the differential equations that arise are very stiff in nature \citep{nejadCOMPARISONSTIFFODE, bovinoASTROCHEMICALMODELLINGPractical2023}, due to
the chemical timescales varying by magnitudes, a stiff ODE solver is needed. A reliable choice is \texttt{DVODE}, which is well tested
for astrochemical problems \citep{garrodComplexChemistryStarForming2008, grassiKROMEPackageEmbed2014, ruaudGasGrainChemical2016}. 

We use \texttt{DVODE} with the Backward Difference Formula (BDF), the finite difference Jacobian approximation mode and
the assumption that the Jacobian is dense. 
More modern alternatives exist, such as \texttt{DLSODES} \citep{hindmarsh1983odepack} and modern BDF Julia solvers \citep{rackauckasDifferentialEquationsjlPerformantFeatureRich2017}.
However, experimentation with the core solver routine is labor intensive
and has a limited return on investment if successful \citep{nejadCOMPARISONSTIFFODE}, and is thus beyond the scope of this
manuscript.

Experiments with a sparse finite difference approximation of the 
Jacobian did not result in a computational speed up either. This is due to the fact that, even though the Jacobian of the gas-phase chemistry
and all other reactions is sparse (<10\%), the introduction of the geometry of having both a surface and bulk drastically
reduces the sparsity to only 40\% as shown in \Cref{fig:jacobian}. The lower number of function evaluations given the sparse pattern does not
outweigh the cost of using the less computationally efficient sparse library. 
The sorting of species by phase first, then by their respective weight, is motivated by creating a band structured jacobian that the solver can effectively simulate. The chosen sorting algorithm and criteria can have a large influence on the ability and effectiveness of the solver \citep{motoyamaEfficientAlgorithmAstrochemical2024}. Further experimentation with preconditioners as provided in \texttt{DLSODES} as well as sorting based on the incidence and stochiometric matrix, might improve performance even further.

The evolution of the chemistry to the next timestep is determined in
the time sampling routine in each of the individual models. The solver will then determine
itself which internal time step size it needs to take to achieve a solution that
satisfies the tolerances and what the order of the solver should be. It
then approximates the Jacobian at the start of every physics timestep using finite
differences. After trying for at  most 10000 internal timesteps, it will
 either return the solution at the target time upon successful integration, or
supply  one of the following five error states:
\begin{itemize}
    \item ISTATE -1: Too many steps need to be taken to reach the target time with the 
    set tolerances and system dynamics. \texttt{UCLCHEM} will divide the target timestep by 10 and try again. This
    can be especially useful when the physics and chemistry are on tightly coupled timescales;
    since \texttt{UCLCHEM} will then update the physics at each of the intermediate target times.
    \item ISTATE -2: The tolerances are too small. \texttt{UCLCHEM} will then try to increase the absolute tolerances
    by a factor of 10. 
    \item ISTATE -3: This is a general error that indicates the solver encountered a non-recoverable
    error. \texttt{UCLCHEM} will fail and raise the error.
    \item ISTATE -4: Means that there was an internal timestep where the error tests did not pass,
    the integration was otherwise successful. \texttt{UCLCHEM} addresses this by dividing the timestep by 10 and
    trying to integrate again.
    \item ISTATE -5: Means that there were internal timesteps where there was no convergence. \texttt{UCLCHEM}
    will divide the target timestep by 10 and try integrating again.
\end{itemize}
With these heuristics, \texttt{UCLCHEM} is able to integrate most astrochemical models successfully.

We also expose the underlying solver statistics to the user, which is essential to
understand how DVODE came to the solution, even more so when a model is struggling to converge. 
The interface exposes the following solver statistics to the user, listed in \Cref{app:solverstats}, allowing the user to identify the culprit when the integration is slow or failing.

It is important to note that the computational
time and complexity of \texttt{UCLCHEM} can vary greatly
as a function of temperature as well as density as seen in \Cref{fig:stiffness}. This highlights that both
ice and gas phase dominated models at low and high temperatures
can be evaluated relatively quickly, but at temperatures
where the competition between freeze-out and desorption becomes important,
the computational time increases because the timestep 
needed to accurately resolve the dynamics becomes much smaller.
This is highlighted by the fact that the largest eigenvalue 
of the ODE peaks at $T=100\;K$, which requires the smallest timestep
and hence the longest wall-time to evaluate.

\subsection{The modern Python interface}
As mentioned before, \texttt{UCLCHEM} now comes
with a much improved Python interface, which relies 
on several different models which can be edited, simulated
and composed together. The most important model
classes are named \texttt{Cloud, ProtostellarCore, CShock, JShock, Collapse} and \texttt{Postprocess}.  
A minimal working example of a Cloud model can be found in
\Cref{app:codeexample}. A full up-to-date documentation
of all models and their current programming interface can be found online \url{uclchem.github.io}. Besides
configuring and running these models directly, they can now
also be used for both sequential and parallel processing.

Naive sequential modelling is achieved easily by running the 
first model, and then passing the first into the second model
as a parameter. The framework will then automatically pass
through the final abundances as starting abundances.

Alternatively, the \texttt{SequentialRunner} was introduced
to pre-define a series of models, which are then executed by
the runner. This can also be combined with the newly introduced 
\texttt{GridRunner}. 
This 
runner can be given the type(s) of model(s) to run, and then be passed 
a parameter dictionary as would be used for 
a singular model, but with some of the parameters having lists of values rather than individual values. The \texttt{GridRunner} 
then uses Python multiprocessing in order to run the 
requested models on the grid. Each model is then saved to a centralised 
\texttt{hdf5} file, set by the user, and becomes accessible once the 
grid finishes. This allows users to quickly define and run 
grids of models.

After (successfully) running a model, \texttt{UCLCHEM} can return the results to the
user, both interactively in Python and as a file on the disk. The most important result is
the array containing the physical conditions as a function of time, as well as
the abundances. 

\begin{figure}
    \centering
    \includegraphics[width=0.95\linewidth]{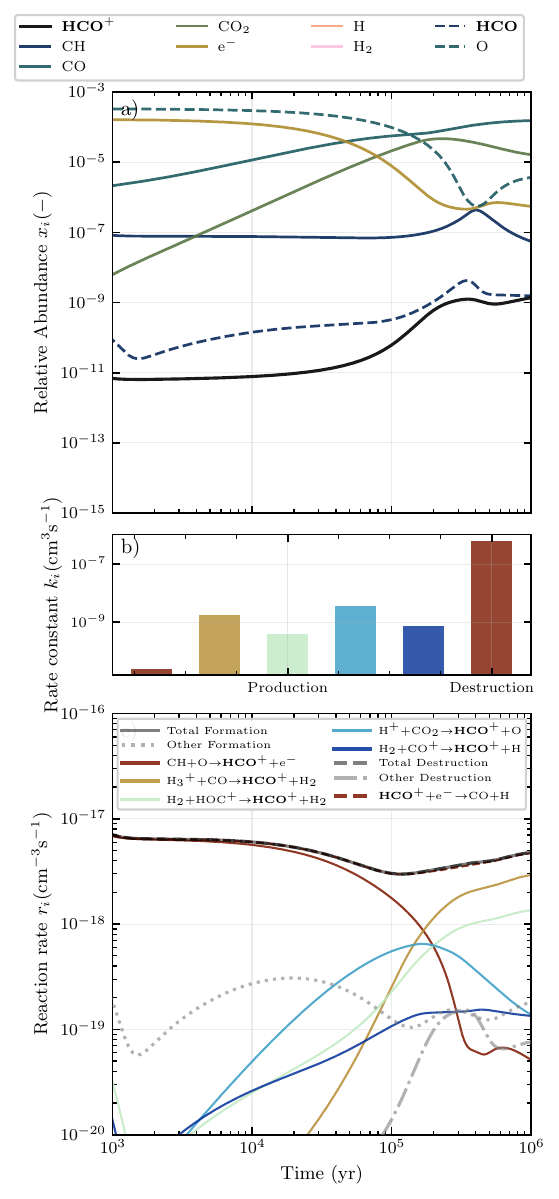}
    \caption{A constant density $n_\mathrm{H}=10^4\;\mathrm{cm}^{-3}$ and temperature $T=75$~K \texttt{UCLCHEM} model. a) shows the evolution of species related to \ch{HCO+}, b) shows the rate constants $k_i$ of
    the most contributing reactions between 1000yr and 1Myr. c) shows the effective reaction rate $r_i$
    including the total formation and destruction as a function of time.} 
    \label{fig:hcoplus_rates}
\end{figure}

The second output lists  both the rate constants $k_j$ and rates $r_j$ of each reaction $j$ ($f_j=\frac{\mathrm{d}x_i}{\mathrm{d}t}|_j$), both of which are useful for understanding the evolution of
the chemistry over time. The reaction rates also include additional correction terms at
the end of the array to account for the geometric effects of the bulk
increasing and decreasing in size as described in \Cref{sec:surfbulkinteraction}.  An example of the evolution
of \ch{HCO^+} and related species for a static model is shown in
\Cref{fig:hcoplus_rates}. As this is a static model (i.e. has a constant temperature),
the rate constants $k_i$ are constant, but due to the large
changes in the abundances of each of the species, the
dominant production mechanism can change from 
$\ch{CH + O -> HCO^+ + e^-}$
at early times to
$\ch{H3^+ + CO -> HCO^+ + H2}$ at later times. This 
once again highlights the nonlinearity of astrochemistry,
with different reaction pathways dominating as the chemical
state evolves over time.

Another important consideration for modern research is the
data storage format. This needs to both be easy to interact with
at runtime and provide an efficient disk storage format,
as is required on high performance computing infrastructure.
The legacy format, while easy to read, is not storage 
efficient nor is it fast to read into 
Python, compared to other methods. 
Because of this, we chose to 
move to storing models into \texttt{hdf5} files using \texttt{h5py} \citep{collettehdf5}. To 
leverage this with the object oriented modelling approach, we 
store all of the attributes of the models in an xarray dataset 
\citep{hoyerXarray2025, hoyerXarrayNDLabeled2017} at runtime. 
Once this dataset is written to disk, it can be loaded back
into \texttt{UCLCHEM} and the user can again interact 
with the model identically as to before the saving.
In addition, multiple models can 
be stored in the same file, allowing users to efficiently 
store several models and share them with other users.

\section{Conclusions and discussion}
\label{sec:conclusion}
In this work, we have presented the \texttt{UCLCHEM} framework, describing
the physics and chemistry it implements, the numerics behind its solver, and
the programming interface through which it is used. \Cref{sec:chem} detailed
the gas and ice-grain chemistry and the mechanisms that move species between
the surface, bulk and gas phases; \Cref{sec:phys} covered the physical
parametrizations available to the user, including cloud collapse,
protostellar warm-up, C- and J-shocks, cosmic ray and radiative transfer
treatments, and the decoupled gas-dust heating and cooling network.
\Cref{sec:uclchem} described how the networks are processed with
\texttt{Makerates} and the considerations of building the inner Fortran routines as well as the new modern Object-Oriented Python
interface. This all allows \texttt{UCLCHEM} to be run interactively and more easily be embedded in
statistical inference and sensitivity-analysis pipelines. Together, these
components make \texttt{UCLCHEM} a flexible framework spanning astrophysical
environments. All code, documentation, tutorials and references are available at
\url{https://uclchem.github.io/}.

\subsection{Limitations and regimes requiring caution}
\texttt{UCLCHEM} couples many different components, each with its own level
of abstraction in the physics, chemistry and numerics, and consequently its
accuracy has limits, typically towards the end of parameter spaces.
While the framework performs well in the regimes highlighted throughout this
paper, it is not always straightforward to state a priori where its results
should be trusted and where they should be treated with additional caution.
We outline the main sources of this uncertainty below.
\begin{itemize}
\item
\textbf{Chemical network.} Not all species and reactions in the network are
equally well constrained \citep{grassiComplexityReductionAstrochemical2012,dijkhuisSensitivityAnalysisInterstellar, vandesandeSenseSensitivityUncertainty2026}. Key, well-investigated molecules are the subject
of ongoing research and their rate coefficients are relatively well known,
whereas rates for less-studied species can carry much larger uncertainties, 
with sometimes parts of the possible reaction network missing altogether.
\citet{dutkowskaCuriousCaseHCOp2026} illustrate this with an in-depth
analysis of the differences between the 2012 and 2022 releases of the UMIST
network and their effect on the ion \ch{HCO^+}: such updates can be confined
to a small region of the physical parameter space, yet still change
predicted abundances by several orders of magnitude within it.
\item
\textbf{Numerical stiffness at high density.} At higher densities the ice
chemistry timescales become very short relative to the gas-phase and
dynamical timescales, which makes the resulting system of differential
equations increasingly stiff. This raises the computational cost of the
integration and, in some cases, may require the user to adjust solver
tolerances or timestepping to obtain a fully converged solution. This
is in addition to the high uncertainty that is still attributed
to the reaction mechanisms on the ices \citep{cuppenLaboratoryComputationalStudies2024a} but also their respective
rates \citep{holdshipBayesianInferenceRates2018, heylExploitingNetworkTopology2020}.
\item
\textbf{Low-density regimes.} Conversely, at lower
densities where radiation increasingly dominates the physical and chemical
state of the gas \citep{rolligPhotonDominatedRegion2007}, \texttt{UCLCHEM} integrates with ease, but its treatment of
radiative transfer and photon-dominated chemistry is less detailed than that
of dedicated PDR codes, and has not been benchmarked against them to the same
extent as its shock and collapse modules. Users targeting these regimes
should be aware of this comparative gap and interpret abundances accordingly.
\end{itemize}
These caveats are not exhaustive, but they capture the regimes where
\texttt{UCLCHEM} fails to provide sound results, and the inherent uncertainty
in the regimes we do model correctly.
Improving within these regimes, alongside
continued benchmarking against sibling gas-grain \citep{jimenez-serraModellingMethanolHydride2025}, PDR \citep{rolligPhotonDominatedRegion2007} and disk chemistry
codes \citep{holdshipChemulatorFastAccurate2021}, remains an ongoing effort for the development of the framework. During these efforts, special attention should be
paid to the overlapping regimes of different codes.

\section*{Acknowledgements}
We thank the reviewers for providing insightful comments that improved
the quality of this manuscript. We thank prof.dr. Herma Cuppen for suggestions that improved the clarity of this manuscript.
We thank all those that provided feedback and requested features for our simulation package. We 
thank Dr Jon Holdship and Dr Ross O'Donoghue for their many contributions to the codebase.
This work is part of a project that has received funding from the European Research Council (ERC) under the European Union’s Horizon 2020 research and innovation programme MOPPEX 833460. TMD is financially supported by the Dutch Astrochemistry Network of the Dutch Research Council (NWO) under grant no.~ASTRO.JWST.001. 
~
\section*{Data Availability} 
The data underlying this article are available in Zenodo at \url{https://doi.org/10.5281/zenodo.15710676} for the source code and \url{https://doi.org/10.5281/zenodo.20747006} for the
code to reproduce the figures from this manuscript.

\bibliographystyle{rasti}
\bibliography{citations}

\appendix

\section{Fitting parameter for grain assisted recombination}
The parameters for grain assisted recombination can be found in \Cref{tab:gar}.
\label{app:gar}
\begin{table*}
\caption{The table with the grain assisted recombination coefficients used in UCLCHEM reproduced from Table 2 of \citet{weingartnerElectronIonRecombinationGrains2001}.}
\label{tab:gar}
\begin{tabular}{llllllll}
\toprule
Ion & $C_0$ & $C_1$ & $C_2$ & $C_3$ & $C_4$ & $C_5$ & $C_6$ \\
\midrule
$\mathrm{H}^{+}$ & 12.25 & $8.074 \times 10^{-6}$ & 1.378 & $5.087\times 10^{2}$ & $1.586 \times 10^{-2}$ & 0.4723 & $1.102 \times 10^{-5}$ \\
$\mathrm{He}^{+}$ & 5.572 & $3.185 \times 10^{-7}$ & 1.512 & $5.115\times 10^{3}$ & $3.903 \times 10^{-7}$ & 0.4956 & $5.494 \times 10^{-7}$ \\
$\mathrm{C}^{+}$ & 45.58 & $6.089 \times 10^{-3}$ & 1.128 & $4.331\times 10^2$ & $4.845 \times 10^{-2}$ & 0.8120 & $1.333 \times 10^{-4}$ \\
$\mathrm{Na}^{+}$ & 2.178 & $1.732 \times 10^{-7}$ & 2.133 & $1.029\times 10^4$ & $1.859 \times 10^{-6}$ & 1.0341 & $3.223 \times 10^{-5}$ \\
$\mathrm{M g}^{+}$ & 2.510 & $8.116 \times 10^{-8}$ & 1.864 & $6.170\times 10^4$ & $2.169 \times 10^{-6}$ & 0.9605 & $7.232 \times 10^{-5}$ \\
$\mathrm{Si}^{+}$ & 2.166 & $5.678 \times 10^{-8}$ & 1.874 & $4.375\times 10^4$ & $1.635 \times 10^{-6}$ & 0.8964 & $7.538 \times 10^{-5}$ \\
$\mathbf{S}^{+}$ & 3.064 & $7.769 \times 10^{-5}$ & 1.319 & $1.087\times 10^2$ & $3.475 \times 10^{-1}$ & 0.4790 & $4.689 \times 10^{-2}$ \\
$\mathrm{K}^{+}$ & 1.596 & $1.907 \times 10^{-7}$ & 2.123 & $8.138\times 10^3$ & $1.530 \times 10^{-5}$ & 1.0380 & $4.550 \times 10^{-5}$ \\
$\mathrm{Ca}^{+}$ & 1.636 & $8.208 \times 10^{-9}$ & 2.289 & $1.254\times 10^5$ & $1.349 \times 10^{-9}$ & 1.1506 & $7.204 \times 10^{-4}$ \\
$\mathrm{Mn}^{+}$ & 2.029 & $1.433 \times 10^{-6}$ & 1.673 & $1.403\times 10^4$ & $1.865 \times 10^{-6}$ & 0.9358 & $4.339 \times 10^{-9}$ \\
$\mathrm{Fe}^{+}$ & 1.701 & $9.554 \times 10^{-8}$ & 1.851 & $5.763\times 10^4$ & $4.116 \times 10^{-8}$ & 0.9456 & $2.198 \times 10^{-5}$ \\
$\mathrm{Ca}^{++} $ & 8.270 & $2.051 \times 10^{-4}$ & 1.252 & $1.590\times 10^2$ & $6.072 \times 10^{-2}$ & 0.5980 & $4.497 \times 10^{-7}$ \\
\bottomrule
\end{tabular}
\end{table*}
\section{\ch{H2} formation on the grains}
\label{app:h2form}
\begin{equation}
R_{\mathrm{H_2,CT}} = \frac{1}{2} v_{\mathrm{therm}}\left[\sigma_{\mathrm{sil}} \xi_{\mathrm{sil}} + \sigma_{\mathrm{gr}} \xi_{\mathrm{gr}}\right]S
\end{equation}
where $\sigma$ is the cross section (cm$^2$, per H nucleus), $\xi$ is the formation efficiency,
thermal velocity is defined as :
\begin{equation}
v_{\mathrm{therm}} = 1.45\times10^5\sqrt{\frac{T_{\mathrm{dust}}}{100\;\mathrm K}}\;\mathrm{cm\,s^{-1}}
\end{equation}
and lastly the sticking coefficient is taken from \citet{hollenbachMoleculeFormationInfrared1979}:
\begin{equation}
S = \left(1 + 0.04\sqrt{T_{\rm{gas}}+T_{\rm{dust}}} + 0.2\left(\frac{T_{\rm{gas}}}{100\;\mathrm K}\right) + 0.08\left(\frac{T_{\rm{gas}}}{100\;\mathrm K}\right)^{2}\right)^{-1}.
\end{equation}

The cross sections for both silicate and graphite are taken to be:
\begin{equation}
\xi = \frac{\varepsilon}{1 + F_1 + F_2}
\end{equation}
where:
\begin{align}
F_1 &= \frac{\mu F}{2\nu_{\mathrm{H_2}} \exp({-E_{\mathrm{H_2}}/T_{\rm dust}})}\\
F_2 &= \frac{1}{4}\left(1 + \sqrt{\frac{E_{\rm HC}-E_{\rm S}}{E_{\rm HP}-E_{\rm S}}}\right)^{2} \exp({-E_{\rm S}/T_{\rm dust})}\\
\varepsilon &= \frac{1}{1 + \dfrac{\nu_{\rm HC}}{2F}\exp\left({\frac{-1.5 E_{\rm HC}}{T_{\rm dust}}}\right)\left(1 + \sqrt{\dfrac{E_{\mathrm{HC}}-E_{\rm S}}{E_{\rm HP}-E_{\rm S}}}\right)^{2}}
\end{align}
where $F = 10^{-10}$~monolayers~s$^{-1}$ is a fixed H flux. The fraction
of new formed $\ch{H2}$ on the grain is $\mu=5\times10^{-3}$, the
saddle point energy between physisorbed and chemisorbed $E_\mathrm{S}$,
$E_{\ch{H2}}$ is the \ch{H2} desorption energy, $E_\mathrm{HP}$ is the physisorbed
binding energy of $\ch{H}$, $E_{\mathrm{HC}}=3\times 10^{4}$~K the chemisorbed binding
energy, $\nu_{\ch{H2}}=3\times 10^{12}$~s$^{-1}$ and $\nu_{\mathrm{HC}}=1.3\times 10^{13}$~s$^{-1}$  the vibrational frequency of a
$\ch{H2}$ surface site and $\ch{H}$ chemisorbed site respectively; $E_{\rm S}$, $E_{\mathrm{H_2}}$, $E_{\mathrm{HP}}$ and $E_\mathrm{{HC}}$ are all in K. Individual
values for silicate and graphite can be found in \Cref{tab:h2form}.

\begin{table}
\centering
\caption{Grain surface parameters for $\ch{H2}$ formation efficiency from \citet{cazauxMolecularHydrogenFormation2002, cazauxH2FormationGrain2004}.}
\label{tab:h2form}
\begin{tabular}{llll}
\toprule
Symbol & Silicate & Graphite & Unit \\
\midrule
$E_{\mathrm{S}}$                    & $110$                 & $260$                 & K \\
$E_{\mathrm{H_2}}$       & $320$                 & $520$                 & K \\
$E_{\mathrm{HP}}$                 & $450$                 & $800$                 & K \\
$E_{\mathrm{HC}}$                 & $3\times10^4$         & $3\times10^4$         & K \\
\bottomrule
\end{tabular}
\end{table}
\section{Table of reaction enthalpies}
A list with reaction enthalpies that are included by default directly are shown in \Cref{tab:clavel} \citep{clavelChemicalThermalEquilibrium1978}.
\label{app:clavel}
\begin{table*}
\centering
\caption{Heating reactions and their
reaction enthalpies.}
\label{tab:clavel}
\begin{tabular}{l c @{\hspace{3em}} l c @{\hspace{3em}} l c}
\toprule
\textbf{Reaction} & \textbf{Energy (eV)} & \textbf{Reaction} & \textbf{Energy (eV)} & \textbf{Reaction} & \textbf{Energy (eV)} \\
\midrule
\ch{H2+ + e- -> H + H} & $10.9$ & \ch{H2 + He+ -> H+ + H + He} & $6.5$ & \ch{H2O + H+ -> H2O+ + H} & $1.0$ \\
\ch{H3+ + e- -> H2 + H} & $9.2$ & \ch{H2 + H2+ -> H3+ + H} & $1.71$ & \ch{H2O + C+ -> HCO+ + H} & $5.3$ \\
\ch{H3+ + e- -> H + H + H} & $4.8$ & \ch{H3+ + O -> OH+ + H2} & $1.64$ & \ch{CH + H+ -> CH+ + H} & $3.0$ \\
\ch{CH+ + e- -> C + H} & $7.2$ & \ch{OH+ + H2 -> H2O+ + H} & $1.2$ & \ch{CH2 + H+ -> CH2+ + H} & $3.2$ \\
\ch{CH2+ + e- -> CH + H} & $6.0$ & \ch{H2O+ + H2 -> H3O+ + H} & $0.8$ & \ch{CH2 + C+ -> C2H+ + H} & $2.4$ \\
\ch{CH3+ + e- -> CH + H + H} & $0.65$ & \ch{H3+ + C -> CH2+ + H} & $2.2$ & \ch{O + OH -> O2 + H} & $7.7$ \\
\ch{CH3+ + e- -> CH2 + H} & $5.0$ & \ch{CH+ + H2 -> CH2+ + H} & $0.14$ & \ch{O2 + H+ -> O2+ + H} & $1.5$ \\
\ch{OH+ + e- -> O + H} & $8.8$ & \ch{CH2+ + H2 -> CH3+ + H} & $0.89$ & \ch{O2 + He+ -> O2+ + He} & $12.5$ \\
\ch{H2O+ + e- -> OH + H} & $7.5$ & \ch{CH+ + O -> CO + H+} & $4.7$ & \ch{O2 + He+ -> O+ + O + He} & $5.9$ \\
\ch{H3O+ + e- -> OH + H + H} & $2.2$ & \ch{OH + C+ -> CO + H+} & $4.4$ & \ch{O2 + C+ -> CO+ + O} & $3.2$ \\
\ch{H3O+ + e- -> H2O + H} & $7.3$ & \ch{OH + C -> CO + H} & $6.7$ & \ch{H3O+ + C -> HCO+ + H2} & $1.5$ \\
\ch{O2+ + e- -> O + O} & $6.9$ & \ch{CO + He+ -> C+ + O + He} & $2.2$ & \ch{CO+ + H2 -> HCO+ + H} & $1.9$ \\
\ch{CO+ + e- -> C + O} & $2.9$ & \ch{CO + H3+ -> HCO+ + H2} & $1.7$ & \ch{CO+ + H -> CO + H+} & $0.4$ \\
\ch{HCO+ + e- -> CO + H} & $7.6$ & \ch{OH + H+ -> OH+ + H} & $0.4$ &  &  \\
\bottomrule
\end{tabular}
\end{table*}

\section{Reaction mechanism summary}
A summary of each reaction mechanism can be found 
in \Cref{tab:ratestable}.
\begin{table*}
\caption{All reaction mechanisms covered within \texttt{UCLCHEM}.}
\label{tab:ratestable}
\begin{tabular}{@{}llll@{}}
\toprule
\textbf{Reaction type} & \textbf{Rate constant $k$} \\ \midrule
Arrhenius-Kooij        &$\alpha(T_{300})^\beta \exp(-\gamma/T)$\\
CR proton              &$\{1,\eta_{ice}\} \alpha \zeta$\\
CR induced photon      &$\{1,\eta_{ice}\} \alpha(T_{300})^\beta\frac{E}{1-\omega}\zeta$\\
UV photon              &$\{1,\eta_{ice}\}\alpha F_\mathrm{UV} \exp(-\gamma A_\mathrm{V})$\\
Freeze                 &$S_i\left(1 + \beta_i \frac{1.671\times10^{-3}}{a_{\mathrm{grain}} T_{\mathrm{gas}}}\right)\sigma_\mathrm{grain}\sqrt{\frac{8k_\mathrm{B}T_\mathrm{gas}}{\pi m_i}}$          \\
Thermal Desorption                 &$\nu_{\mathrm{des}}^i\exp\left(-E_{\mathrm{bind}}^i / T_\mathrm{dust}\right)$                \\
\ch{H2} formation & See \Cref{app:h2form} above $T=150$ K, $\texttt{\{LH,ER\}DES}$ below                             \\
CR Desorption          &$4\pi\zeta \langle \pi a_\mathrm{grain}^2 n_\mathrm{grain} \rangle$\\
CR induced photon Desorption       &$\langle \pi a_\mathrm{grain}^2 n_\mathrm{grain} \rangle Y F_\mathrm{P} \zeta$            \\
ISRF Desorption        &$\langle \pi a_\mathrm{grain}^2 n_\mathrm{grain} \rangle Y F_\mathrm{P} \frac{F_\mathrm{UV}}{\eta} \exp\left(-1.8 A_\mathrm{V}\right)$ \\
Eley-Rideal                     & $k_{\mathrm{freeze},j} P_{\mathrm{reac},ij} n_s(j) n_s(i)/n_s$                              \\
Langmuir-Hinshelwood                     &$\frac{f_{\mathrm{comp}}\left(k_{\mathrm{diff},i}+k_{\mathrm{diff},j}\right)}{N_\mathrm{sites}n_\mathrm{d}} n_s(i) n_s(j)$    \\
Encounter Desorption & $\frac{2k_{\mathrm{diff}}^{\ch{H2}}}{N_\mathrm{sites}n_\mathrm{dust}}\frac{k_{\text{des}}^{\text{H}_2\text{ on H}_2}}{k_{\text{des}}^{\text{H}_2\text{ on H}_2} + k_{\text{diff}}^{\text{H}_2\text{ on H}_2}}$ &  \\

Ion polar       &   $\begin{aligned}k_1&=\alpha \beta (0.62 + 0.4767 \gamma T_{300}^{-1/2}\\ k_2&=\alpha \beta (1 + 0.0967 \gamma T_{300}^{-1/2} + \frac{\gamma ^2}{10.526} T_{300}^{-1}\end{aligned}$     \\
Cosmic Rays            &     $F_\mathrm{CR} \alpha \beta (\gamma/100) \zeta  $             \\
Solid Excitation       &  $\alpha \frac{1}{1.8\times 10^{-8} N_\mathrm{sites}}\left(\sqrt{E_{\rm bind}/m}+\sqrt{E_{\rm bind}/m}\right)$       \\
Relaxation of excited species      &   $\nu_{\mathrm{des}}^i$             \\
Bulk to surface swapping     &  $ \sqrt{E_b/m}\exp\left(-E_b / T_\mathrm{dust}\right)$\\
Grain-assisted recombination & $\frac{0.6\times10^{-14}{C_0}}{1+C_1\psi^{C_2}\left(1+C_3 T_\mathrm{gas}^{C_4}\psi^{-C_5-C_5\ln T_\mathrm{gas}}\right)}, \psi=F_\mathrm{UV}\sqrt{T_\mathrm{gas}} / n_e$ & \\

\bottomrule
\end{tabular}
\end{table*}

\section{Exposed solver statistics}
\label{app:solverstats}
By default, \texttt{UCLCHEM} exposes the following
solver statistics from DVODE to the user:
\begin{itemize}
    \item \textbf{T} Start time in years
    \item \textbf{HU} The step size in years
    \item \textbf{HCUR} The next step size in years
    \item \textbf{TCUR} The current time (end of the latest step) in years
    \item \textbf{TOLSF} The tolerance scale factor
    \item \textbf{NST} The number of steps taken so far
    \item \textbf{NFE} The number of function evaluations ($\dot{Y}(n_i,T, F_\mathrm{UV}, \zeta)$)
    \item \textbf{NJE} The number of jacobian evaluations so far
    \item \textbf{NQU} The order used for the last step
    \item \textbf{NQCUR} The order to attempt on the next step
    \item \textbf{IMXER} The index of the largest magnitude contribution to the error vector when ISTATE=-4 or ISTATE=-5
    \item \textbf{NLU} The number of LU decompositions
    \item \textbf{NNI} The number of nonlinear Newton iterations used so far
    \item \textbf{NCFN} The number of convergence failures of the nonlinear solver
    \item \textbf{NETF} The number of error test failures for the integrator.
\end{itemize}

\section{A bird's-eye view of \texttt{UCLCHEM}}
In \Cref{fig:birdseyeview}, a flowchart is provided that highlights the
code and data flow of \texttt{uclchem}. 
\begin{figure*}
\centering
\resizebox{\linewidth}{!}{%
\begin{tikzpicture}[
  font=\sffamily\small,
  node distance=8mm and 6mm,
  every node/.style={align=center},
  box/.style={draw, rounded corners=2pt, minimum width=4.85cm, minimum height=1.0cm,
              line width=0.7pt, inner sep=7pt, text width=4.4cm},
  py/.style={box, fill=pyfill, draw=pyline},
  fo/.style={box, fill=fofill, draw=foline},
  art/.style={box, fill=artfill, draw=artline},
  core/.style={fo, line width=1.2pt},
  arr/.style={-{Latex[length=2mm,width=1.6mm]}, line width=0.8pt, draw=black!65,
              rounded corners=2pt},
  arrret/.style={-{Latex[length=2mm,width=1.6mm]}, line width=0.7pt, draw=black!45,
                 dashed, rounded corners=2pt},
  arrstate/.style={{Latex[length=1.8mm,width=1.4mm]}-{Latex[length=1.8mm,width=1.4mm]},
                 line width=0.6pt, draw=foline, densely dotted, rounded corners=2pt},
  lbl/.style={font=\scriptsize, text=black!55},
  phase/.style={font=\scriptsize\bfseries\sffamily, text=black!55, anchor=north west,
                xshift=4pt, yshift=-4pt}
]

\node[py] (data) {\textbf{User defined reaction\\network and configuration}
  \subtext{\texttt{Makerates/data/\{species, \ldots, reactions\}.csv} \\ \texttt{user\_settings.yaml}}};
\node[py, right=22.2mm of data] (makerates) {\textbf{Makerates}\subtext{Makerates/Makerates.py\\\texttt{uclchem-makerates} (CLI)\\\texttt{uclchem.makerates}}
  \note{Construct chemical network and check network and settings}};
\node[art, right=22.2mm of makerates] (netf90) {\textbf{Fortran network files}
  \subtext{\texttt{\{network, odes, f2py\_constants\}.f90}}
  \note{Encode the chemical network as arrays in the binary}};

\draw[arr] (data) -- (makerates) node[lbl, midway, above] {reads};
\draw[arr] (makerates) -- (netf90) node[lbl, midway, above] {writes};

\coordinate (bandA-head) at ($(data.north)+(0,0.62)$);
\begin{scope}[on background layer]
\node[fit=(data)(makerates)(netf90)(bandA-head), inner sep=8pt, fill=bandfill, draw=bandline, rounded corners=4pt] (bandA) {};
\end{scope}
\node[phase] at (bandA.north west) {1~~BUILD THE NETWORK \quad\normalfont\scriptsize\itshape\mdseries user curates the chemical network};

\node[fo, minimum width=3.75cm, text width=3.25cm, minimum height=4.4cm,
      below=28mm of netf90, xshift=-4.67cm] (chem)
  {\textbf{Chemistry modules}
  \subtext{\texttt{\{chemistry, rates, surfacereactions, photoreactions\}.f90}}
  \note{Functions that relate to core gas and ice chemistry}};
\node[fo, minimum width=3.75cm, text width=3.25cm, minimum height=4.4cm,
      right=5mm of chem] (phys)
  {\textbf{Physics modules}
  \subtext{\texttt{\{physics-core, heating, coolant\_module, extinction\}.f90}}
  \note{+ modular functions: \texttt{\{cloud, collapse, hotcore, cshock, jshock, postprocess\}\_mod.f90} \\
  Functions that relate to visual extinction, radiation, density and temperature}};

\coordinate (chemPhysMid) at ($(chem)!0.5!(phys)$);
\node[core, minimum width=4.9cm, text width=4.4cm, below=30mm of chemPhysMid] (wrap)
  {\textbf{\texttt{wrap.f90}}\\[1pt]
  {the main routine}\\[4pt]
  {\texttt{solveAbundances} is the core routine that solves the chemistry and physics together, as a function of time. \\Exposed via functional f2py API in Python:}\\[2pt]
  {\texttt{\{cloud, collapse, hot\_core, cshock, jshock, postprocess\}(param\_dict, ...)}}};

\node[fo, minimum width=2.0cm, text width=1.6cm, right=4mm of wrap, yshift=-2.6cm] (dvode)
  {\textbf{\texttt{dvode.f90}}\subtext{stiff ODE integrator}};

\node[art, below=25mm of wrap] (ext)
  {\textbf{\texttt{uclchemwrap.so}}
  \subtext{compiled Python extension}
  \note{solves, then fills the pre-allocated arrays in place (intent(in,out))}};

\coordinate (ncBendRef) at ($(netf90.south)!0.4!(chem.north)$);
\coordinate (nc1) at (netf90.south |- ncBendRef);
\coordinate (ncOutW) at ($(chem.west)+(-0.35,0)$);
\coordinate (nc2) at (ncOutW |- ncBendRef);
\coordinate (nc3) at (ncOutW |- chem.west);
\draw[arr] (netf90.south) -- (nc1) -- (nc2) -- (nc3) -- (chem.west);
\node[lbl, anchor=south] at ($(nc1)!0.5!(nc2)+(0,0.1)$) {compiles into};

\coordinate (turnChemW) at ($(wrap.north)+(-1.1,0)$);
\coordinate (cwBendRef) at ($(chem.south)!0.5!(turnChemW)$);
\coordinate (cw1) at (chem.south |- cwBendRef);
\coordinate (cw2) at (turnChemW |- cwBendRef);
\draw[arr] (chem.south) -- (cw1) -- (cw2) -- (turnChemW);
\node[lbl, anchor=west] at ($(chem.south)+(0.08,-0.16)$) {includes};

\coordinate (turnPhysW) at ($(wrap.north)+(1.1,0)$);
\coordinate (pwBendRef) at ($(phys.south)!0.5!(turnPhysW)$);
\coordinate (pw1) at (phys.south |- pwBendRef);
\coordinate (pw2) at (turnPhysW |- pwBendRef);
\draw[arr] (phys.south) -- (pw1) -- (pw2) -- (turnPhysW);
\node[lbl, anchor=west] at ($(phys.south)+(0.08,-0.16)$) {includes};

\coordinate (dvTurn) at (dvode.north |- wrap.east);
\draw[arr] (dvode.north) -- (dvTurn) -- (wrap.east);
\node[lbl, anchor=south] at ($(dvTurn)!0.5!(wrap.east)+(0,0.05)$) {integrates};
\draw[arr] (wrap) -- (ext) node[lbl, midway, right] {compiles into};

\coordinate (bandB-head1) at ($(chem.north)+(0,0.62)$);
\coordinate (bandB-head2) at ($(phys.north)+(0,0.62)$);
\begin{scope}[on background layer]
\node[fit=(chem)(phys)(wrap)(dvode)(ext)(bandB-head1)(bandB-head2), inner sep=8pt, fill=bandfill, draw=bandline, rounded corners=4pt] (bandB) {};
\end{scope}
\node[phase] at (bandB.north west) {2~~COMPILE \quad\normalfont\scriptsize\mdseries\texttt{pip install .}\ to compile and install};

\coordinate (dataMakMid) at ($(data)!0.5!(makerates)$);
\node[py, below=37.6mm of dataMakMid, xshift=-2.40cm, text width=4.4cm] (model)
  {\textbf{uclchem.model}\note{pick one of six physics models}\\[4pt]
  {\ttfamily\selectfont
    \begin{tabular}{@{}l@{\hspace{7mm}}l@{}}
      Cloud & Collapse\\
      ProtostellarCore & CShock\\
      JShock & Postprocess\\
    \end{tabular}}};

\node[py, minimum width=3.3cm, text width=2.95cm, minimum height=4.9cm,
      below=11mm of model, xshift=-1.9cm] (arrays)
  {\textbf{Simulation arrays} \texttt{physics\_array, chemical\_abun\_array, \dots}
  \note{user can access via Python API: \texttt{model.get\_dataframes()}}};
\node[py, minimum width=3.3cm, text width=2.95cm, minimum height=4.9cm,
      right=5mm of arrays] (pdict)
  {\textbf{User runtime options}
  \texttt{param\_dict = \{\\
  \ \ "initialDens": 1e5,\\
  \ \ "initialTemp": 27,\\
  \ \ "finalTime": 3e6\}}
  \note{The user can specify runtime specific parameters for each simulation:\\ \texttt{Model(param\_dict, \ldots)}}};

\node[py, below=7.5mm of arrays, xshift=1.9cm] (analysis)
  {\textbf{\texttt{uclchem.\{analysis, plot\}}}\subtext{reaction pathways, rate/abundance plots}
  \note{user can easily visualize using:\\
  \texttt{model.create\_abundance\_plot(\\{}["CO"])}}};

\node[py, below=2mm of analysis] (adv)
  {\textbf{\texttt{uclchem.advanced}}\subtext{\texttt{HeatingSettings, NetworkState, GeneralSettings}}
  \note{reads and edits live Fortran module state directly }};

\coordinate (mpBendRef) at ($(model.south)!0.5!(pdict.north)$);
\coordinate (mp1) at (model.south |- mpBendRef);
\coordinate (mp2) at (pdict.north |- mpBendRef);
\draw[arr] (model.south) -- (mp1) -- (mp2) -- (pdict.north);
\node[lbl, anchor=south] at ($(mp1)!0.5!(mp2)+(0,0.04)$) {builds};

\coordinate (raBendRef) at ($(arrays.south)!0.5!(analysis.north)$);
\coordinate (ra1) at (arrays.south |- raBendRef);
\coordinate (ra2) at (analysis.north |- raBendRef);
\draw[arr] (arrays.south) -- (ra1) -- (ra2) -- (analysis.north);
\node[lbl, anchor=south west] at ($(ra1)+(0.10,0.05)$) {read after simulation};

\coordinate (callTurn) at ($(pdict.east)+(2.5,0.12)$);
\coordinate (callBotRef) at ($(ext.west)+(0,0.9)$);
\coordinate (callVertBot) at (callTurn |- callBotRef);
\draw[arr] ($(pdict.east)+(0,0.12)$) -- (callTurn) |- (callBotRef);
\node[lbl, anchor=south] at ($(pdict.east)!0.5!(callTurn)+(0,0.06)$) {calls via f2py};

\coordinate (arrOutW) at ($(arrays.west)+(-0.18,0)$);
\coordinate (arrGapY) at ($(analysis.south)!0.5!(adv.north)$);
\coordinate (arrGapL) at (arrOutW |- arrGapY);
\coordinate (arrGapR) at ($(adv.east)+(1.3,0)$);
\coordinate (arrGapRPt) at (arrGapR |- arrGapY);
\coordinate (arrBotY) at ($(adv.south)+(1.0,+1.0)$);
\coordinate (arrBotPt) at (arrGapR |- arrBotY);
\coordinate (arrTarget) at ($(ext.west)+(0,0)$);
\coordinate (arrRisePt) at (arrGapR |- arrTarget);
\draw[arrstate] (arrays.west) -- (arrOutW) -- (arrGapL) -- (arrGapRPt) -- (arrBotPt) -- (arrRisePt) -- (arrTarget);

\coordinate (advTarget) at ($(ext.west)+(0,-0.55)$);
\coordinate (advOut) at ($(adv.east)+(1.0,0)$);
\coordinate (advVertBot) at (advOut |- advTarget);
\draw[arrstate] (adv.east) -- (advOut) |- (advTarget);

\coordinate (bandC-head) at ($(model.north)+(0,0.62)$);
\begin{scope}[on background layer]
\node[fit=(model)(pdict)(arrays)(analysis)(adv)(bandC-head), inner sep=8pt, fill=bandfill, draw=bandline, rounded corners=4pt] (bandC) {};
\end{scope}
\node[phase] at (bandC.north west) {3~~RUN A MODEL \quad\normalfont\scriptsize\itshape\mdseries every simulation};

\coordinate (interBandGap) at ($(bandC.east)!0.5!(bandB.west)$);
\node[lbl, anchor=south] at ($(interBandGap |- arrTarget)+(0,0.05)$) {filled in place};
\node[lbl, anchor=south] at ($(interBandGap |- advTarget)+(0,0.05)$) {module state access};

\begin{scope}[on background layer]
\node[fit=(bandA)(bandB)(bandC), inner sep=0pt] (bandAll) {};
\end{scope}
\node[py, minimum width=0.32cm, minimum height=0.32cm, text width=0pt, inner sep=0,
      anchor=west] (l1) at ($(bandAll.west |- bandC.south)+(0.3,-0.9)$) {};
\node[right=1.5mm of l1, anchor=west, font=\scriptsize] {Python};
\node[fo, minimum width=0.32cm, minimum height=0.32cm, text width=0pt, inner sep=0,
      right=2.5cm of l1, anchor=west] (l2) {};
\node[right=1.5mm of l2, anchor=west, font=\scriptsize] {Fortran (compiled)};
\node[art, minimum width=0.32cm, minimum height=0.32cm, text width=0pt, inner sep=0,
      right=3.0cm of l2, anchor=west] (l3) {};
\node[right=1.5mm of l3, anchor=west, font=\scriptsize] {generated / build artifact};

\end{tikzpicture}%
}
\caption{A diagram highlighting the inner workings of the Fortran code, Python code, configuration files and
data flow of \texttt{UCLCHEM}}
\label{fig:birdseyeview}

\end{figure*}

\section{Code example}
\label{app:codeexample}
The following code demonstrates initialization and execution of a single-point isothermal cloud model:

\begin{lstlisting}[language=Python, caption={Initialize and run a cloud model}]
import uclchem

param_dict = {
    "endAtFinalDensity": False,
    "freefall": False,
    "initialDens": 1e4,
    "initialTemp": 10.0,
    "finalTime": 1.0e6,
    "rout": 0.1,
    "baseAv": 1.0,
}
cloud = uclchem.model.Cloud(
    param_dict=param_dict,
)
\end{lstlisting}

Then verify the simulation ran successfully and the elemental abundances are conserved:

\begin{lstlisting}[language=Python, caption={Check convergence and elemental conservation}]
cloud.check_error()

df = cloud.get_dataframes()

cloud.check_conservation(element_list=["H", "N", "C", "O", "S"])
\end{lstlisting}

Then easily plot the gas and ice abundances of a few
species of interest:

\begin{lstlisting}[language=Python, caption={Generate abundance plot}]
fig, ax = cloud.create_abundance_plot(
    species=["H", "H2", "$H", "$H2", "H2O", "$H2O",
             "CO", "$CO", "$CH3OH", "CH3OH"],
    figsize=(10, 7),
)
ax.set(xscale="log", ylim=(1e-15, 1), xlim=(1e3, 1e6))
\end{lstlisting}

Species prefixed with \texttt{\$} denote total ice-phase abundance.

\bsp	%
\label{lastpage}
\end{document}